\documentclass[a4paper,11pt]{article}
\pdfoutput=1 

\usepackage{jheppub} 

\usepackage[T1]{fontenc} 
\usepackage{blindtext}
\usepackage{epstopdf}
\usepackage{graphicx}
\usepackage{epsfig}
\usepackage{dcolumn}  
\usepackage{bm}    
\usepackage{caption}
\usepackage{subcaption}
\usepackage{amssymb} 
\usepackage{tikz}
\usetikzlibrary{arrows}
\usepackage{nccmath}
\usepackage{xcolor}
\usepackage{epstopdf}
\usepackage{graphicx,wrapfig,lipsum}
\usepackage{epsfig}
\usepackage{amsmath,bm}
\usepackage{amsfonts}  
\usepackage{amsmath}  
\usepackage{slashed}  
\usepackage{enumitem}
\usepackage[mathscr]{euscript}
\usepackage{tabu}
\usepackage{epsfig}
\makeatletter
\g@addto@macro\bfseries{\boldmath}
\makeatother

\def\l1{{{1-loop}}}

\def\n1{\Bigg|_{n=1}}

\def\n{{(n)}}

\usepackage[T1]{fontenc} 
\usepackage{tikz}
\usepackage{amsmath,amssymb}
\usepackage{relsize}
\usepackage{latexsym}
\usepackage{leftidx}
\usepackage{xcolor}
\usepackage{diagbox}
\usepackage[T1]{fontenc}
\usepackage{array}
\usepackage{makecell}
\usepackage{csquotes}
\usepackage{tikz}
\usepackage{enumitem}
\usepackage{setspace}
\usepackage{multirow}
\usepackage{amsmath,amssymb}
\usepackage{relsize}
\usepackage{latexsym}
\usepackage{leftidx}
\usepackage{xcolor}
\usepackage{csquotes}
\usepackage{tikz}
\usetikzlibrary{decorations.pathmorphing}
\usetikzlibrary{arrows.meta}
\usepackage{enumitem}
\usetikzlibrary{decorations.markings}
\usetikzlibrary{decorations.pathmorphing}
\usetikzlibrary{decorations.markings}
\usetikzlibrary{decorations.pathmorphing}
\usepackage{pifont}
\usepackage{hyperref}
\usepackage{url}

\usepackage{bookmark}

\title{\textbf{\textsf{Large $N$ Wess-Zumino model at finite temperature and large chemical potential in $3d$
}}}
\author{Srijan Kumar}
\affiliation{\vspace{.1cm} Centre for High Energy Physics, \\ Indian Institute of Science,\\
	C. V. Raman Avenue, Bangalore 560012, India.}
\emailAdd{srijankumar@iisc.ac.in}
\abstract{ We consider the  supersymmetric Wess-Zumino model at large $N$  in $(2+1)$ dimension. We introduce a chemical potential($\mu$)  at finite temperature($T$). The non-trivial fixed point of this model is described by a pair of coupled gap equations. This fixed point behaves as a thermal CFT for all values of the coupling. 
	We find that at large chemical potential these coupled equations simplify and solutions become analytically tractable. 
	We solve them analytically for all values of the coupling  at this limit.
	The	solutions admit a systematic series expansion in $\frac{T}{\mu}$.
	 Thus, using the solutions of the gap equation at large chemical potential  we can evaluate the analytic form of the partition function, stress tensor and spin-1 current as a perturbative expansion in orders of $\frac{T}{\mu}$.
	Applying the OPE inversion formula on the scalar and fermion two point functions of the theory, we compute higher spin currents at large $\mu.$ 
	
}
\begin{document}
	\maketitle
	\flushbottom
	\section{Introduction}
	The 	Wess-Zumino model was introduced as one of the simplest examples of supersymmetric quantum field theory in the literature \cite{Wess:1974tw}, see for review \cite{Moshe:2003xn}.  The model consists of equal number of bosonic and fermionic degrees of freedom and it allows both intra-species as well as inter-species interactions. The two kinds of couplings are governed by a single coupling constant.  Thermodynamics of this model in $(2+1)$ dimensions was studied in \cite{DeWolfe:2019etx} at large $N$ and at zero chemical potential. The theory behaves as a thermal CFT for any value of the coupling constant and the conformal fixed point is characterised by the bosonic and fermion thermal masses satisfying a pair of coupled gap equations. And the  thermodynamic quantities e.g. pressure, energy density had been evaluated by solving the coupled gap equations for the thermal masses numerically.\\

	We introduce a chemical potential $\mu$ in this model starting from the supersymmetric Wess-Zumino model without chemical potential in the fundamental of $O(2N)$. Rewriting this in terms of complex scalars and fermions we get the $U(N)$ invariant action with  complex scalars and Dirac fermions in dimension $d=3$ with the Wess-Zumino interaction. Now the action can be linearised using the Hubbard-Stratanovich trick introducing  auxiliary fields at large $N$. Only the zero mode of the auxiliary field contributes at the leading order. Finally we evaluate the partition function at the saddle point of the path integral in the zero mode of the auxiliary fields. Two saddle point conditions give the pair of gap equations involving the chemical potential $\mu$.  Now the modifications in the gap equations caused by the presence of chemical potential $\mu$ is simple to understand. The pair of coupled gap equation describing the conformal fixed point of the theory at any value of the coupling constant $\lambda$ is given by
	\begin{align}\label{gap eqn intro}
		\frac{m_f}{4\lambda}&=-\frac{m_b}{2\pi}-\frac{T}{2\pi}[\log(1-e^{\frac{\mu-m_b}{T}})+\log(1-e^{-\frac{m_b+\mu}{T}})],\nonumber\\
		\frac{m_{b}^2}{8 \lambda}&=\frac{3 m_f^2}{8\lambda}+\frac{m_f^2}{2\pi}+\frac{m_f T}{2\pi}[\log(1+e^{\frac{\mu-m_f}{T}})+\log(1+e^{-\frac{m_f+\mu}{T}})],
	\end{align}
	where 	$m_b$ and $m_f $ are thermal masses of bosons and fermions respectively,  $T$ is the temperature. \\
	
	We observe that at large chemical potential $\frac{\mu}{T}\to \infty$, the gap equations \eqref{gap eqn intro} simplifies and can be solved at this limit. The solutions admit systematic series expansion in $\frac{T}{\mu}$.
	At the leading order in large $\mu$ in units of temperature($T$), the solutions of the gap equation  for the thermal mass of scalars and fermions are given as
	\begin{align}\label{mb mf}
		m_b=\mu,\qquad
		m_f=	\frac{(\sqrt{4 \lambda ^2+3 \pi ^2}-2 \lambda ) \mu }{3 \pi }.
	\end{align}

	Study of thermal one point functions of CFT operators has emerged as an important question in recent developments in the context of holography as well as finite temperature CFTs. In this paragraph, we provide a summary of the recent advancements in this field. It has been shown that holographic CFT thermal one point functions can probe the interior geometry of the $AdS$ black hole in the bulk \cite{Grinberg:2020fdj,Berenstein:2022nlj,David:2022nfn,Singhi:2024sdr}. Evaluating thermal one point function from field theory turns out to be a challenging task. In this context a class of large $N$ vector models serve as a suitable examples of CFT's at finite temperature where thermal one point functions of higher spin currents are computed based on the simplification at large $N$.  Methods to evaluate thermal one point functions of arbitrary spin($l$) currents has been developed for $O(N)$ model at large $N$ examining the OPE of the two point functions  at the non-trivial  fixed point \cite{Sachdev:1993pr,Chubukov:1993aau} of the theory, the method is called OPE inversion formula \cite{Iliesiu:2018fao}. At zero temperature the model in $d=3$ is dual to the higher spin gravity theory in the  bulk $AdS_4$ geometry \cite{Klebanov:2002ja}.  The OPE inversion formula has been generalised for large $N$ Gross-Neveu model for fermions \cite{Petkou:2018ynm,David:2023uya}. Thermal one point functions of higher spin currents in presence of chemical potential in the theory are computed by applying the OPE inversion formula on the two point functions of twisted fields in \cite{Karydas:2023ufs, David:2024naf}. For these large $N$ vector models the leading $\frac{1}{N}$ corrections to the free energy as well as the thermal one point functions are discussed in \cite{Diatlyk:2023msc}. Bootstrap techniques are also developed to evaluate thermal OPE coefficients of the $O(N)$ model containing the thermal one point functions of higher spin currents at finite $N$ \cite{Barrat:2024fwq}. We have computed thermal one  point functions of CFT higher spin currents for $O(N) $ model on $S^1\times S^2$  in \cite{David:2024pir} as a perturbative expansion at large radius of the sphere $S^2$. These one point functions of higher spin currents show interesting behaviour both at large spin($l$) and large space time dimension($d$) \cite{David:2023uya,David:2024naf}.  For any general CFT on $S^1\times S^2$ the one point functions are examined in \cite{Buric:2024kxo}.	\\

	In this paper, we will compute thermal one point functions of the operators constructed out of scalar and fermion bilinears for the large $N$ Wess-Zuminio model with chemical potential $\mu$ in $3d$ applying the OPE inversion formula. Finally, incorporating the solutions of the gap equations at large $\mu$ given in \eqref{mb mf}, in these thermal one point functions we obtain their large $\mu$ behaviour.  We study thermal one point functions of the classes of operators in this Wess-Zumino model at the large $N$ conformal fixed point given as follows.
	\begin{align}\label{0,+,-,O}
		\langle{\cal O}[n,l]\rangle&\equiv \langle\phi^* D_{\mu_1}\cdots D_{\mu_l}D^{2n}\phi\rangle\equiv b_{\cal O}[n,l]T^{\Delta_{\cal O}}\nonumber,\\
	\langle	\tilde{{ {\cal O}}}_{0}[n,l]\rangle&\equiv
		\langle\psi_\alpha^\dagger D_{\mu_1}\cdots D_{\mu_l}D^{2n}\psi_\alpha\rangle\equiv b_{{\cal \tilde O}_0}[n,l] T^{\Delta_{{\cal \tilde O}_0}}\nonumber,\\
	\langle	\tilde	 {{ {\cal O}}}_{+}[n,l]\rangle&\equiv \langle\psi_\alpha^\dagger\gamma^{\alpha\beta}_{\mu_1}
		 D_{\mu_2}\cdots D_{\mu_l}D^{2n}\psi_\beta\rangle\equiv b_{{\cal O}_+}[n,l] T^{\Delta_{{\cal \tilde O}_-}} \nonumber,\\
		\langle\tilde	  {{ {\cal O}}}_{-}[n,l]\rangle&\equiv\langle \psi_\alpha^\dagger\gamma^{\alpha\beta}_{\mu}D^{\mu}
		D_{\mu_1}\cdots D_{\mu_l}D^{2n}\psi_\beta\rangle\equiv b_{{\cal O}_-}[n,l]T^{\Delta_{{\cal \tilde O}_+}}.
	\end{align}
	where $D_\nu=\partial_\nu+\mu\delta_{\nu,0}$ with $\nu=0,1,2$ and $\gamma_\mu^{\alpha\beta}$ denotes the Dirac gamma matrices.
	These bilinear operators in the above four equations are understood to be traceless symmetric parts of the objects written above. $\Delta_{{\cal O}},\Delta_{{\cal O}_0},\Delta_{{\cal O}_+},\Delta_{{\cal O}_-}$ are the scaling dimensions of the corresponding operators.  \\
	
	Notation: Throughout this paper the notation $\tilde{\cal O}$  stands for the operators constructed of the fermion bilinears while $\cal O$   denotes scalar bilinears.\\
	
	The thermal expectation value of these operators given in \eqref{0,+,-,O} are evaluated by studying the OPEs of the thermal two point functions of complex scalars and fermions simultaneously with the use of OPE inversion formula.
	We allow the fermion bilinear to occur in the OPE of the scalar two point correlator and vice versa. The reason is that  the interactions between scalar and fermion fields in the action may cause non-trivial three point function coefficient between scalar, scalar, fermionic bilinear operator and also fermion, fermion, scalar bilinear operators.   Starting from such a general structure of these OPEs we see that such mixing of operators are not present at the leading order in large $N$. The three-point function between fermion, fermion, and scalar bilinear comes out to be zero from the OPE inversion formula directly. Although for scalar, scalar, fermion bilinear it does not come out directly from the OPE inversion formula, we argue that this three point function is zero based on the fact that it is known to vanish at the Gaussian fixed point as well as at the non-trivial fixed point obtained at strong coupling. We rigorously show that it indeed vanishes for $l=1$ and $2$.  Once we obtain the thermal one point functions in terms of the thermal masses of bosons and fermions we take the large $\mu$ limit of the results with the use of the large $\mu$ solutions of the gap equations \eqref{mb mf}. At large $\mu $ we get the thermal one point function of the scalar bilinear ${\cal O}[0,l]$ as a systematic series expansion as given below
	\begin{align}
		\lim_{\mu   \rightarrow \infty} a_{\cal O}[0, l ] =
		\frac{\mu^{l+1} \Gamma (l+1)}{2^{2-l}\pi  \Gamma (2 l+1)} \Bigg[ 1+\frac{\sqrt{4\lambda^2+3\pi^2}-2\lambda}{6\lambda} +& \sum_{n =1}^l \frac{   (l - n +1)_{2n} \zeta(n+1)  }{ 2^n n!} \frac{T^{n+1} }{ |\mu|^{n+1} }\Bigg] .
	\end{align}
	The corrections appear as $O(e^{-\frac{\mu}{T}})$.
	The above expression also contains overall constant factors along with $\langle {\cal O}[0,l]\rangle$ due to the OPE coefficient which does not depend on the dynamics at finite temperature.
	And at large $\mu$ the fermion bilinear operator ${\cal O}_+[0,l] $ has thermal expectation given by the following expansion in terms of regularised hypergeometric functions
	\begin{align}\label{asym for fer intro}
		&\lim_{\mu\to\infty}	a_{{\cal\tilde O}_+}[0,l]=	-\frac{l (c \mu )^{l+1} }{\pi  2^{l+1} }
		\sum _{k=0}^{l+1} \frac{(2 \pi i  )^{k}B_k\left(\frac{1}{2}\right) }{\left(\frac{1}{2}\right)_l(2c)^{k}}  \Big(\frac{T}{\mu}\Big)^k
		\frac{ \, _2\tilde{F}_1\left(k-l-1,k+l;k;\frac{c-1}{2 c}\right)}{k! (k+l)_{2-2k}}	,\nonumber\\
		&    {\rm where }\qquad  c=\frac{(\sqrt{4 \lambda ^2+3 \pi ^2}-2 \lambda )  }{3 \pi }.
	\end{align}
	This expression also has an overall constant factor due to the OPE coefficients independent of finite temperature dynamics. Note that this expression is always real as $B_k(\frac{1}{2})$ vanishes for $k$ being odd. Again, the corrections occur in  order $O(e^{-\frac{\mu}{T}})$.
	\\
	
	The organisation of the paper is as follows. In the section \ref{sec 2} we discuss general structures of the OPEs for scalar and fermion two point functions for theories with scalar-fermion interactions. The section \ref{sec 3} contains the description of the Wess-Zumino model at finite chemical potential and its analytic solutions at large $N$ and large chemical potential. Section \ref{4} has discussions of the OPE inversion formula and the results for the thermal one point functions.  Section \ref{sec 5} has the large $\mu$ asymptotic formula for the thermal one point functions. The agreement of the spin-1 and spin-2 current with the partition function calculation is discussed in section \ref{sec 6}. The appendix \ref{A} has the details of regularisation used to evaluate the partition function for free scalar and free fermions.
	
		\section{OPEs in theories with both bosons and fermions}\label{sec 2}
		In this section we introduce the operator product expansion of the two point correlation functions of scalars and fermions. It is essential to examine both of these while studying the models constituted of both bosonic and fermion fields. Important and non-trivial modifications arise due to the inter-species interaction between scalar and fermionic fields, which allows the possibility of  bosonic two point function decomposing into fermionic bilinear operators in the OPE and as well as the two point correlator of fermion fields admitting the presence of the bosonic bilinear operators in its OPE. Identifying all sorts operators present in the OPEs of both scalar and fermion two point functions, we organise these OPEs in terms of Gegenbauer polynomials. This sets the stage for the computation of the thermal one point functions for all theses operators constructed from scalar and fermion bilinears by the application of OPE inversion formula at finite temperature.
		\subsection{Scalar two point function}
		We consider theories with both bosons and fermions and allow interaction between them.
		The general structure of OPEs for the two point function of scalars $\phi$ may contain fermionic bilinears owing to such interactions as three point function of boson, boson and fermionic bilinears can be non-vanishing in general.
		\begin{align}\label{OPE boson}
			\phi^*(x)\phi(0)=\sum_\mathcal{O} C_{\phi \phi \mathcal{O}}|x|^{-2\Delta_\phi+\Delta_\mathcal{O}-l} x_{\mu_1}\cdots x_{\mu_l} \mathcal{O}^{\mu_1\cdots\mu_l}\nonumber\\
			+\sum_\mathcal{\tilde O} C_{\phi\phi\gamma^{\alpha\beta}{\cal\tilde O}_{\alpha\beta}}|x|^{-2\Delta_\phi+\Delta_{\tilde{\mathcal{O}}}-l-1} x_{\mu_1}\cdots x_{\mu_l} x_\mu \gamma^{\alpha\beta\,\mu}\tilde{\mathcal{O}}^{\mu_1\cdots\mu_l}_{\alpha\beta}\nonumber\\+
			\sum_\mathcal{\tilde O} C_{\phi\phi{\cal\tilde O}_{\alpha\alpha}}|x|^{-2\Delta_\phi+\Delta_{\tilde{\mathcal{O}}}-l} x_{\mu_1}\cdots x_{\mu_l}  \tilde{\mathcal{O}}^{\mu_1\cdots\mu_l}_{\alpha\alpha},
		\end{align}
		$\mathcal{O} $ and $ O_{\alpha\beta}$ refer to the bosonic and fermionic bilinear  operators respectively.
		$\alpha, \beta$ stand for the spinor indices due to the fermionic fields and $\mu_1,\cdots,\mu_l$ are space-time indices, repeated indices are understood to be summed over. $C_{\phi\phi{\cal O}}, C_{\phi\phi\gamma^{\alpha\beta}{\cal O}_{\alpha\beta}}$ and $C_{\phi\phi{\cal O}_{\alpha\alpha}}$ are the relevant OPE coefficients which are the ratio of the structure constant to the normalisation of the operators appearing in the OPE. The bosonic and fermionic bilinear operators arising in the above OPE have the schematic forms as given below,
		\begin{align}
			\mathcal{O}^{\mu_1\cdots\mu_l}\equiv \phi^*\partial_{\mu_1}\cdots\partial_{\mu_l}(\partial^2)^n\phi\nonumber,\\
			\tilde{	\mathcal{ O}}_{\alpha\beta}^{\mu_1\cdots\mu_l}\equiv \psi_\alpha^\dagger\partial_{\mu_1}\cdots\partial_{\mu_l}(\partial^2)^n\psi_\beta.
		\end{align}
		
		Now the spatial rotational and spatial translational invariance of the thermal vacuum fix the structure of the thermal one point functions of bosonic bilinears except a numerical constant to be determined. Thus thermal one point functions of the class of bosonic operators denoted by ${\cal O}$ can be represented as traceless symmetric tensors as given below,
		\begin{align}
			\langle \mathcal{O}_{\mu_1\cdots\mu_l}\rangle=b_\mathcal{O}T^\Delta (e_{\mu_1} e_{\mu_2}\cdots e_{\mu_l}-\text{traces}),
		\end{align}
		
		$e_{\mu}$'s are unit vectors along imaginary time direction $\tau$. $b_\mathcal{O} $ is the numerical constant specific to a theory.
		And using the following mathematical identity,
		\begin{align}\label{identity}
			|x|^{-l}( x^{\mu_1}\cdots x^{\mu_l})( e_{\mu_1}\cdots e_{\mu_l}-\text{traces})=\frac{l!}{2^{l}(\nu)_{l}}C_{l}^{(\nu)}(\eta),
		\end{align}
		where $C_l^{\nu} (\eta) $ is Gegenbauer polynomial with
		\begin{align}
			\eta=\frac{\tau}{|x|}, \qquad {\rm and} \qquad \nu=\frac{d-2}{2}.
		\end{align}
		The 1st term from \eqref{OPE boson} is expressed in terms of Gegenbauer polynomials,

		\begin{align}\label{1st term}
			\sum_\mathcal{O} C_{\phi \phi \mathcal{O}}|x|^{-2\Delta_\phi+\Delta_\mathcal{O}-l} x_{\mu_1}\cdots x_{\mu_l} \mathcal{O}^{\mu_1\cdots\mu_l}=	\sum_\mathcal{O} a_{\cal O}^b|x|^{-2\Delta_\phi+\Delta_\mathcal{O}}C_{l}^{(\nu)}(\eta),
		\end{align}
		where,
		\begin{align}
			a_{\cal O}^b=C_{\phi \phi \mathcal{O}}\frac{b_{\mathcal{O}}l!}{2^{l}(\nu)_{l}}T^{\Delta_{\cal O}}.
		\end{align}
		Here the notation \enquote{$b$} in the superscript means that it is derived from the OPE for the bosonic 2-point function. This is used to specify the kind of structure constant multiplied with the one point functions.\\
		
		Now we focus on the 2nd term from the OPE \eqref{OPE boson}, and the fermionic bilinear operator appearing in  this term is decomposed into a symmetric part, trace part and an antisymmetric part,
		as was done in \cite{David:2023uya}.

		\begin{eqnarray} \label{trace}
			\gamma^{\nu \; \alpha\beta} {\cal  \tilde O}^{ \mu_1\cdots \mu_l}_{ \beta\alpha} &=&
			\left[ \frac{1}{l +1} \left(  \gamma^{\nu \; \alpha\beta} {\cal \tilde O}^{ \mu_1\cdots \mu_l}_{ \beta\alpha}  + 
			\gamma^{\mu_1 \; \alpha\beta} {\cal \tilde O}^{ \mu_2\cdots \mu_l\nu}_{ \beta\alpha} + {\rm cyclic}  \right) 
			-{\rm Traces}  \right] + {\rm Traces} \nonumber \\ 
			&& +  \frac{1}{l +1}  \left(  \gamma^{\nu \; \alpha\beta} {\cal \tilde O}^{ \mu_1\cdots \mu_l}_{ \beta\alpha} - 
			\gamma^{\mu_1 \; \alpha\beta} {\cal \tilde O}^{ \nu \mu_2 \cdots \mu_l}_{ \beta\alpha}  \right) 
			+ \cdots ( l-1 ){\rm terms}   ,
		\end{eqnarray}
		The `Traces' stands for
		\begin{equation}
			{\rm Traces} = \frac{2}{(l+1)(d + l-1)} 
			\left( \delta^{\nu\mu_1} \gamma^{\rho\;\alpha\beta} {\cal O}^{ \rho{\mu_2} \cdots \mu_l}_{ \beta\alpha} 
			+ \delta^{\nu\mu_2} \gamma^{\rho\;\alpha\beta} {\cal O}^{ \rho{\mu_3} \cdots \mu_l \mu_1 }_{ \beta\alpha} 
			+ {\rm cyclic} \right) .
		\end{equation}
		Finally, using the mathematical identity \eqref{identity} after substituting this tensor structure \eqref{trace}, the 2nd term from the OPE \eqref{OPE boson} is simplified to,
		\begin{align} \label{2nd term}
			&\sum_\mathcal{\tilde O} C_{\phi\phi\gamma^{\alpha\beta}{\cal O}_{\alpha\beta}}|x|^{-2\Delta_\phi+\Delta_{\tilde{\mathcal{O}}}-l-1} x_{\mu_1}\cdots x_{\mu_l} x_\mu \gamma^{\alpha\beta\,\mu}{\mathcal{O}}^{\mu_1\cdots\mu_l}_{\alpha\beta}\nonumber\\&=
			\sum_{\tilde {\cal O}  } |x|^{\Delta_{{\cal\tilde O}} - 2\Delta_{\phi} } 
			\left(  a_{{\cal\tilde O}_+} C^{(\nu)}_{l+1} (\eta)  + \frac{ 2l}{(l+1)( d+ l-1)}  a_{{\cal \tilde O}_{-} } C^{(\nu)}_{l-1}(\eta)  \right),
		\end{align}
		where,
		\begin{eqnarray}
			a_{{ \cal\tilde  O}_{+}}^b = b_{{\cal O}_+} T^{\Delta_{{\cal O}_+}}
			\frac{(l+1)! }{2^{l+1} (\nu)_{l+1} } C_{\phi\phi\gamma^{\alpha\beta}{\cal O}_{\alpha\beta}},
			\\ \nonumber
			a_{{\cal \tilde O}_{-}}^b = b_{ {\cal O}_- } T^{\Delta_{{\cal O}_-} }
			\frac{(l-1)!}{2^{l-1} (\nu)_{l-1} } C_{\phi\phi\gamma^{\alpha\beta}{\cal O}_{\alpha\beta}}.
		\end{eqnarray}
		Now in the 3rd term in the OPE \eqref{OPE boson} the fermionic operator of the class $\tilde{\cal O}_{0}$ again can be written as traceless symmetric tensor as follows,
		\begin{align}
			\langle	\mathcal{ O}_{\alpha\alpha}^{\mu_1\cdots\mu_l}\rangle_\beta= b_{ {\cal O}_0} T^\Delta (e_{\mu_1} e_{\mu_2}\cdots e_{\mu_l}-\text{traces}),
		\end{align}
		And also the third term from \eqref{OPE boson} is written as,
		\begin{align}\label{3rd term}
			\sum_\mathcal{\tilde O} C_{\phi\phi{\cal \tilde O}_{\alpha\alpha}}|x|^{-2\Delta_\phi+\Delta_{\tilde{\mathcal{O}}}-l} x_{\mu_1}\cdots x_{\mu_l}  {\mathcal{\tilde O}}^{\mu_1\cdots\mu_l}_{\alpha\alpha}=
			\sum_\mathcal{\tilde O}  a_{{\cal\tilde O}_0}^b|x|^{-2\Delta_\phi+\Delta_\mathcal{\tilde O}}C_{l}^{(\nu)}(\eta)\nonumber
		\end{align}
		where
		\begin{align}
			a_{ {\cal\tilde  O}_0}^b=	C_{\phi\phi{\cal \tilde O}_{\alpha\alpha}}\frac{b_{\mathcal{ \tilde O}_0}l!}{2^{l}(\nu)_{l}} .
		\end{align}
		Thus combining all the three terms in the OPE \eqref{OPE boson} as obtained in \eqref{1st term}, \eqref{2nd term} and \eqref{3rd term} respectively, and if the anomalous dimensions of the operators present in the OPE are vanishing at the leading order in large $N$\footnote{Our analysis in the section \ref{4} will establish the fact that for large $N$ Wess-Zumino model in $3d$ operator dimensions do not get corrected at the leading order in large $N$.}, we can use  
		\begin{align}\label{all deltas}
			\Delta_{\cal O}=2\Delta_\phi+2n+l \qquad {\rm and} \qquad \Delta_{\cal \tilde O}=2 \Delta_{\psi}+2n+l\nonumber,\\
			{\rm where}\qquad \Delta_\phi=\frac{d-2}{2}\qquad {\rm  and} \qquad \Delta_{\psi}=\frac{d-1}{2}.
		\end{align}
		And present the OPE \eqref{OPE boson} in terms of the Gegenbauer polynomials as
		\begin{align}
			\langle		\phi(x) \phi(0)\rangle=&\sum_{n,l=0}^\infty a_{\mathcal{O}}^b |x|^{2n+l} C_l ^{\nu}(\eta) 
			+\sum_{l,n=0}^\infty |x|^{2n+l+1} 
			a_{{\cal\tilde O}_+}^b C^{(\nu)}_{l+1} (\eta)  \\
			& +\sum_{l=1,n=0}^\infty |x|^{2n+l+1}  \frac{ 2l }{(l+1)( l+d-1)}  a_{{\cal \tilde O}_{-} }^b C^{(\nu)}_{l-1}(\eta)  
			+\sum_{n,l=0}^\infty  a_{\mathcal{\tilde O}_0}^b |x|^{2n+l+1} C_l^{\nu}(\eta) \nonumber
		\end{align}
		Reorganising the sums in the above equation we obtain the OPE as,
		\begin{align}\label{ph ph OPE}
			\langle		\phi(x) \phi(0)\rangle=\sum_{n=1}^\infty a_{\mathcal{O}}^b[n,0] |x|^{2n} C_0 ^{(\nu)}(\eta) +\sum_{l=1}^\infty (a^b_{\mathcal{O}}[0,l]+ a^b_{{\cal\tilde O}_+}[0,l]) |x|^{l} C_l ^{(\nu)}(\eta) 
			\nonumber\\
			+\sum_{l=0,n=1}^\infty |x|^{2n+l}  \Big(a^b_{\mathcal{O}}[n,l]+ a^b_{{\cal\tilde O}_+}[n,l]+\frac{ 2(l+1)}{(l+2)( d+ l)} a^b_{{\cal\tilde O}_{-} }[n-1,l]\Big) C^{(\nu)}_{l}(\eta) \nonumber \\
			+\sum_{n,l=0}^\infty a^b_{\mathcal{\tilde O}_0}[n,l] |x|^{2n+l+1} C_l^{(\nu)}(\eta).
		\end{align}
		The above structure will be useful in the analysis of the OPE inversion formula in the Section \ref{4}.
		
		\subsection{Fermion two point functions}
		
		Similar to the bosonic case, due to the fermion-boson interaction the OPE of the fermionic two point function allows presence of bosonic bilinear operators as the  three point function of fermion, fermion and bosonic bilinear can be non-vanishing in general just as we had similar considerations for the bosonic OPE. Thus combining all sorts of operators allowed, the general structure for the OPE of the fermionic two point function $\psi_\alpha(x)$ can be written as,
		\begin{align}\label{OPE fer}
			g_{\alpha\beta}(x)=	\langle	\psi^\dagger _\alpha(0)\psi_\beta(x)\rangle=\sum_{\mathcal{\tilde O}} C_{\psi\psi\mathcal{\tilde  O}} |x|^{-2\Delta_{\psi}+\Delta_\mathcal{\tilde O}-l}  x_{\mu_1}\cdots x_{\mu_l} \langle \mathcal{\tilde  O}_{\mu_1\cdots\mu_l}^{\alpha\beta}\rangle\nonumber\\
			+ \sum_{\mathcal{O}} C_{\psi\psi\mathcal{ O}} |x|^{-2\Delta_{\psi}+\Delta_\mathcal{O}-l-1}  x_{\mu_1}\cdots x_{\mu_l} x^\mu \gamma^{\alpha\beta}_\mu\langle \mathcal{ O}_{\mu_1\cdots\mu_l}\rangle \nonumber\\
			+ \sum_{\mathcal{O}} C_{\psi\psi\mathcal{ O}} |x|^{-2\Delta_{\psi}+\Delta_\mathcal{O}-l}  x_{\mu_1}\cdots x_{\mu_l}\delta^{\alpha\beta}\langle \mathcal{ O}_{\mu_1\cdots\mu_l}\rangle .
		\end{align}
		The 1st term in the above operator product expansion of fermion two point correlator has the bilinear operator of fermion with spinor indices $\alpha,\beta$ while the 2nd and 3rd terms are decompositions of the fermion two point function into bosonic bilinears through two different projection channels in the OPE. And $C_{\psi\psi\cal \tilde O},C_{\psi\psi\cal O}$ and $C_{\psi\psi\cal O}$ refer to the relevant OPE coefficients.
		
		Now, we define three different correlators of fermion field $g_1(x),g_2(x), g_3(x)$ obtained by projecting the $g_{\alpha\beta}(x) $ in three different channels as shown below,
		\begin{align}\label{diff ch fer}
			g_1(x)&=\langle \psi_\alpha^\dagger(x)\psi_\alpha(0)\rangle_\beta\nonumber,\\
			g_2(x)&=\frac{1}{|x|}\langle \gamma^{\alpha\beta}_\mu x^\mu\psi_\alpha^\dagger(x)\psi_\beta(0)\rangle_\beta\nonumber,\\
			g_3(x)&=\langle \gamma_\mu^{\alpha\beta}\partial^\mu\psi_\alpha^\dagger(x)\psi_\alpha(0)\rangle_\beta.
		\end{align}
		By studying these three OPE channels simultaneously, using OPE inversion formula the thermal one point functions of different classes of both bosonic and fermionic bilinears can be computed separately.
		\subsection*{The correlator $g_1(x)$}
		Form \eqref{OPE fer} and \eqref{diff ch fer} the OPE for the correlator $g_1(x)$ is given by,
		\begin{align}
			g_1(x)=\langle \psi^\dagger_\alpha(x)\psi_\alpha(0)\rangle&=\sum_{\mathcal{\tilde O}} C_{\psi\psi\mathcal{ \tilde O}} |x|^{-2\Delta_{\psi}+\Delta_\mathcal{\tilde O}-l}  x^{\mu_1}\cdots x^{\mu_l} \langle \mathcal{ \tilde O}_{\mu_1\cdots\mu_l}^{\alpha\alpha}\rangle\nonumber\\
			&\qquad+ 2\sum_{\mathcal{O}} C_{\psi\psi\mathcal{ O}} |x|^{-2\Delta_{\psi}+\Delta_\mathcal{O}-l}  x^{\mu_1}\cdots x^{\mu_l}\langle \mathcal{ O}_{\mu_1\cdots\mu_l}\rangle\\
			&=\sum_\mathcal{ O}a^f_{\mathcal{\tilde O}_0} |x|^{-2\Delta_\psi+\Delta_\mathcal{\tilde O}} C_{l}^{(\nu)}(\eta)
			+2\sum_\mathcal{O} a^f_{\mathcal{ O}} |x|^{-2\Delta_\psi+\Delta_\mathcal{O}} C_{l}^{(\nu)}(\eta),
		\end{align}
		Where 
		\begin{align}\label{structure const fer}
			a^f_{{\cal \tilde O}_0}=C_{\psi\psi \mathcal{\tilde O}}\frac{b_{\mathcal{\tilde O}_0}l!}{2^{l}(\nu)_{l}}T^{\Delta_{\cal O}}\qquad {\rm and} \qquad a^f_{\cal O}=C_{\psi\psi{\cal O}}\frac{b_{\mathcal{ O}}l!}{2^{l}(\nu)_{l}}T^{\Delta_{\cal O}}.
		\end{align}
		Certainly the notation \enquote{$f$} is used to identify the thermal one point functions of the operators derived from the OPE of fermion 2-point correlators.
		Similar to the case of scalars we use \eqref{all deltas} to obtain
		\begin{align}\label{g_1 ope}
			g_1(x)=\sum_{n,l=0}^\infty a_{\mathcal{ \tilde O}_0}^f[n,l] |x|^{2n+l} C_{l}^{(\nu)}(\eta)
			+2\sum_{n,l=0}a^f_{\mathcal{ O}}[n,l] |x|^{2n+l-1} C_{l}^{(\nu)}(\eta).
		\end{align}
		
		\subsection*{The correlator $g_2(x)$}
		Projecting the OPE \eqref{OPE fer} in the vector channel using \eqref{diff ch fer} we can compute  $g_2(x)$ as the following

		\begin{align}\label{g_2}
			g_2(x)=	\langle g_{\alpha\beta} \frac{\gamma^{\alpha\beta}_\nu x^\nu}{|x|}\rangle=
			\sum_{\mathcal{\tilde O}} C_{\psi^\dagger\psi\mathcal{\tilde O}}|x|^{\Delta_\mathcal{\tilde O}-2 \Delta_\psi-l-1}x^{\mu_1}\cdots x^{\mu_l}x^\nu\langle\gamma^{\alpha\beta}_\nu\mathcal{\tilde O}_{\alpha\beta\mu_1\cdots\mu_l}\rangle\nonumber\\
			+\sum_\mathcal{O}C_{\psi^\dagger\psi\mathcal{O}}|x|^{\Delta_\mathcal{O}-2\Delta_\psi-l}\, x^{\mu_1}\cdots x^{\mu_l} \langle \mathcal{O}_{\mu_1\cdots\mu_l}\rangle .
		\end{align}
		The 1st term from the above equation can be organised as the following, using the similar relation as \eqref{2nd term},
		\begin{align}
			&	\sum_{\mathcal{\tilde O}} C_{\psi^\dagger\psi\mathcal{\tilde O}}|x|^{\Delta_\mathcal{\tilde O}-2 \Delta_\psi-l-1}x^{\mu_1}\cdots x^{\mu_l}x^\nu\langle\gamma^{\alpha\beta}_\nu\mathcal{\tilde O}_{\alpha\beta\mu_1\cdots\mu_l}\rangle\nonumber\\
			&	=
			\sum_{{\mathcal {\tilde O}}  } |x|^{\Delta_{\mathcal {\tilde O}} - 2\Delta_{\psi} } 
			\left(  a^f_{\mathcal {\tilde O}_+} C^{(\nu)}_{l+1} (\eta)  + \frac{ 2l}{(l+1)( d+ l-1)} a^f_{\mathcal {\tilde O}_{-} } C^{(\nu)}_{l-1}(\eta)  \right),
		\end{align}
		where 
		\begin{eqnarray}
			a_{{ \cal\tilde  O}_{+}}^f = b_{{\cal O}_+} T^{\Delta_{{\cal O}_+}}
			\frac{(l+1)! }{2^{l+1} (\nu)_{l+1} } C_{\psi\psi{\cal \tilde O}},
			\\ \nonumber
			a_{{\cal \tilde O}_{-}}^f = b_{ {\cal O}_- } T^{\Delta_{{\cal O}_-} }
			\frac{(l-1)!}{2^{l-1} (\nu)_{l-1} } C_{\psi\psi{\cal \tilde O}}.
		\end{eqnarray}
		Thus, substituting this in \eqref{g_2} we have the fermionic correlator $g_2$ given by the following
		\begin{align}
			g_2(x)=\sum_{{\mathcal {\tilde O}}} |x|^{\Delta_{\mathcal {\tilde O}} - 2\Delta_{\psi} } 
			\left(  a^f_{\mathcal {\tilde O}_+} C^{(\nu)}_{l+1} (\eta)  + \frac{ 2l}{(l+1)( d+ l-1)} a^f_{\mathcal {\tilde O}_{-} } C^{(\nu)}_{l-1}(\eta)  \right)\nonumber\\
			+\sum_{\mathcal{O}} |x|^{\Delta_\mathcal{O}-2\Delta_\psi} a^f_\mathcal{O} C_l^{(\nu)}(\eta).
		\end{align}
		The above OPE is reorganised as the following with the use of \eqref{all deltas}
		\begin{align}\label{g2 gen struct}
			g_2(x)&=\frac{a^f_\mathcal{O}[0,0]}{|x|}C_0^{(\nu)}(\eta)+\sum_{l=1}^{\infty} (a^f_{\mathcal{\tilde O}_+}[0,l]+a^f_{\mathcal{O}}[0,l]) |x|^{l-1} C_l^{(\nu)}(\eta)\nonumber\\
			&+\sum_{n=1,l=1}^{\infty} \Big(a^f_{\mathcal{\tilde O}_+}[n,l]+a^f_{\mathcal{O}}[n,l]+\frac{2(l+1)}{(l+2)(d+l)}a^f_{\mathcal{\tilde O}_-}[n-1,l+1]\Big) |x|^{2n+l-1} C_l^{(\nu)}(\eta)
		\end{align}
		\subsection*{The Correlator $ g_3(x)$}
	
		Using the similar approach, from \eqref{OPE fer} and \eqref{diff ch fer}, it is  easy to evaluate the  correlator $g_3(x)$ to be given by the following expression
		\begin{align}
			g_3(x)=&\sum_{\mathcal{\mathcal{O}}} |x|^{\Delta_\mathcal{O}-2\Delta_\psi-1} (\Delta_\mathcal{O}-2\Delta_\psi+2) a_{\mathcal{O}}[n,l] C_l^{(\nu)} (\eta)\nonumber\\
			&+ \sum_{\mathcal{\mathcal{O}}}  |x|^{\Delta_\mathcal{O}-2\Delta_\psi} a_\mathcal{O}[n,l+1] C_{l+1}^{(\nu)}(\eta)\nonumber\\
			&+ \sum_{{\cal \tilde O} \in \psi^\dagger \times \psi }|x|^{ \Delta_{\cal \tilde O} - 2\Delta_{\psi} -1}
			\Big[ ( \Delta_{\cal \tilde O} - 2\Delta_{\psi} - l) a_{{\cal \tilde O}_+} C^{(\nu)}_{l+1} (\eta)    \nonumber\\ 
			& \qquad\qquad\qquad   +\Big( l + \frac{2l ( \Delta_{\cal \tilde O} - 2\Delta_{\psi} - l)}{(l+1)( l+ d-1) } \Big ) a_{{\cal \tilde O}_-}  C^{(\nu)}_{l-1} (\eta) 
			\Big].
		\end{align}
		Now, again by using \eqref{all deltas} in  the above expression 
		we get
		\begin{align}
			&g_3(x)=\sum_{l=1}^\infty 2|x|^{l-2} a^f_\mathcal{O}[0,l] C_l^{(\nu)}(\eta)+\sum_{n=0}^{\infty}\Big(|x|^{2 n}(1+\frac{2n}{d})a^f_{\mathcal{\tilde O}_-}[n,0]+|x|^{2n-2}a^f_{\cal O}[n,0]\Big)C_0^{(\nu)} (\eta)+ \nonumber\\
			&\sum_{\substack{n=0\\l=1}}^{\infty}|x|^{2 n+l}\left[2a^f_\mathcal{O}[n+1,l]+
			2 (n+1) a^f_{\mathcal{\tilde O}_+}[n+1,l]+\bigg(l+1+\frac{4 n(l+1)}{(l+2)(l+d)}\bigg)a^f_{\mathcal{\tilde O}_-}[n,l] \right] C_l^{(\nu)}(\eta).
		\end{align}
		\subsection{Twisted fields}
		For the theories with chemical potential $\mu$, we should look for the one point functions of conserved currents constructed from scalar $\tilde \phi$ and fermion fields $\tilde \psi$ with covariant derivative structure as given below
		\begin{align}\label{4 Os}
			&{\cal O}=\tilde\phi^* D_{\mu_1}\cdots D_{\mu_l}\tilde\phi\nonumber,\\
			&\tilde {\cal O}_0=\tilde\psi^\dagger D_{\mu_1}\cdots D_{\mu_l}\tilde\psi\nonumber,\\
			&\tilde{\cal O}_+=\tilde\psi^\dagger\gamma_{\mu_1}D_{\mu_2}\cdots D_{\mu_l}\tilde\psi\nonumber,\\
			&\tilde {\cal O}_-=\tilde\psi^\dagger\gamma^{\mu}D_{\mu}D_{\mu_1}\cdots D_{\mu_l}\tilde\psi,
		\end{align}
		where $D_{\mu}=\partial_\mu-i\hat\mu\delta_{\tau,0}$ and $\hat\mu=i\mu$.
		Note that we will work with imaginary chemical potential $\mu=-i\hat \mu$ till we actually find thermal one point functions of the operators given above and finally we analytically continue to real chemical potential of the results.\\
		
		Thus if one uses the OPE of the scalar field with periodic boundary condition or fermions with anti-periodic boundary condition it is not easy to read out thermal one point functions of such operators with covariant derivatives from the OPE. But  with the following redefinition of the fields 
		\begin{align}
			\tilde	\phi(x)=e^{i\hat\mu\tau} \phi(x) \qquad {\rm and} \qquad \tilde \psi(x)=e^{i\hat\mu\tau}\psi(x),
		\end{align}
		$\tilde \psi$ and $\tilde \phi$ are called the twisted fields and they satisfy the twisted boundary conditions as given below
		\begin{align}
			\phi(\tau+1,\vec x)=e^{-i\hat\mu} \phi(\tau,\vec x)\qquad {\rm and}\qquad \psi(\tau+1,\vec x)=-e^{-i\hat\mu} \psi(\tau,\vec x).
		\end{align}
		We can see that the operators in \eqref{4 Os} can be written in terms of normal derivatives instead  of covariant derivatives using the twisted fields.
		\begin{align}
			&{\cal O}=\phi^*\partial_{\mu_1}\cdots \partial_{\mu_l}\phi\nonumber,\\
			&\tilde {\cal O}_0=\psi^\dagger\partial_{\mu_1}\cdots \partial_{\mu_l}\psi\nonumber,\\
			&\tilde {\cal O}_+=\psi^\dagger\gamma^{\mu_1}\partial_{\mu_2}\cdots D_{\mu_l}\psi\nonumber,\\
			&\tilde {\cal O}_-=\psi^\dagger\gamma^{\mu}\partial_{\mu}\partial_{\mu_1}\cdots \partial_{\mu_l}\psi.
		\end{align}
		Thus we will study the OPEs of the twisted fields where the operators with the covariant derivative structures are manifestly present. 
		Note that from now onward we use $\tilde \phi$ and $\tilde \psi $ to denote untwisted fields while $ \phi$ and $ \psi $ denote the twisted fields.

		\section{Wess-Zumino model with chemical potential($\mu$) and large $\mu$}
		\label{sec 3}
		The critical Wess-Zumino model is a well-studied example of a CFT at arbitrary coupling strength. The model admits simplifications at large $N$ limit. The non-trivial fixed point of the theory at finite temperature is characterised by the bosonic and fermionic thermal masses denoted by $m_b$ and $m_f$ respectively, which satisfy the large $N$ gap equations or the saddle point equation of the partition function on $S^1\times R^2$. We consider the supersymmetric action \cite{Moshe:2003xn,DeWolfe:2019etx} for the Wess-Zumino model in $(2+1)d$ in the fundamental of $O(2N)$, with $2N$ real scalars and $2N$ real fermions. Now by combining pairs of real scalars and pairs of real fermions as complex scalars and Dirac fermions respectively we obtain the $U(N)$ Wess-Zumino model with $N$ complex scalars and $N$ Dirac fermions in $d=3$.\footnote{We thank Justin David for suggesting this point.}. Now we introduce a chemical potential $\mu$ in this setup. The Euclidean action for $U(N)$ Wess-Zumino model with $N$ complex scalars $\tilde \phi_a$ and fermions $\tilde \psi_a$ with sextic intra-species interaction among the complex scalars and the scalar-fermion interation is described below, 
		\begin{align}\label{WZ action}
			S=\int_0^\beta d\tau\int d^2x \Big[(\partial_\tau-\mu)\tilde\phi_a^*(\partial_\tau+\mu)\tilde\phi_a+{|\partial_i\tilde\phi_a|^2}+ \tilde\psi_a^\dagger\gamma_0(\partial_\tau+\mu)\tilde\psi_a+{\psi_a^\dagger\gamma^i\partial_i\tilde\psi_a}\nonumber\\
			+\frac{64\lambda^2}{N^2}(\tilde\phi_a^*\tilde\phi_a)^3+\frac{\lambda}{N} \tilde\psi_a^\dagger\tilde\psi_a\tilde\phi_c^*\tilde\phi_c+8 \frac{\lambda}{N} \tilde\psi_a^\dagger\tilde\psi_b\tilde\phi^*_b\tilde\phi_a+\frac{8\lambda}{N}(\tilde\psi^\dagger_a\tilde\phi_a+\tilde\psi_a\tilde\phi^*_a)^2\Big],
		\end{align}
		where $a=1,\cdots,N$ and $i=1,2.$
		The partition function is given by the following euclidean path integral on $S^1_\beta\times R^2$,
		\begin{align}
			Z=\int {\cal D}\tilde\phi^*{\cal D}\tilde\phi {\cal D}\tilde\psi^\dagger{\cal D}\tilde\psi e^{-S},
		\end{align} 
		The last term involving $(\psi^\dagger_a\phi_a+\psi_a\phi_a)^2$ in the action \eqref{WZ action} does not contribute to the leading order in large $N$ so we drop this term. Then we apply the standard trick of Hubbard-Stratanovich transformation  to simplify the interaction terms  by introducing auxiliary fields $\sigma$ and $\zeta $ as shown below. The use of the definition for the Dirac delta as given below,
		\begin{align}
			1=\int{\cal D}\zeta {\cal D} \sigma e^{i\int \zeta(\sigma-\frac{\phi_c^*\phi_c}{N})},
		\end{align}
		brings the partition function in the following form,
		\begin{align}
			Z=\int {\cal D}\tilde\phi^*{\cal D}\tilde\phi {\cal D}\tilde\psi^\dagger{\cal D}\tilde\psi {\cal D}\zeta {\cal D} \sigma \exp\Big[-\int d^3x\Big((\partial_\tau-\mu)\tilde\phi_a^*(\partial_\tau+\mu)\tilde\phi_a+{|\partial_i\tilde\phi_a|^2}\nonumber\\+ \tilde\psi_a^\dagger\gamma_0(\partial_\tau+\mu)\tilde\psi_a
			+{\tilde\psi_a^\dagger\gamma^i\partial_i\tilde\psi_a}
			+64\lambda^2N\sigma^3+8\lambda  \sigma\tilde\psi_a^\dagger\tilde\psi_a-i\zeta\sigma+i\zeta\frac{\tilde\phi_a^*\tilde\phi_a}{N}\Big)\Big].
		\end{align}
		Rescaling $\sigma $ and $\zeta$ in the following manner,
		\begin{align}
			\sigma=\frac{\sigma'}{8\lambda}\qquad {\rm and}\qquad \zeta=-iN\zeta'.
		\end{align}
		The partition function becomes,
		\begin{align}\label{part fn}
			Z=\int {\cal D}\tilde\phi^*{\cal D}\tilde\phi {\cal D}\tilde\psi^\dagger{\cal D}\tilde\psi {\cal D}\zeta {\cal D} \sigma \exp\Big[-\int d^3x\Big((\partial_\tau-\mu)\tilde\phi_a^*(\partial_\tau+\mu)\tilde\phi_a+{|\partial_i\tilde\phi_a|^2}\nonumber\\+ \tilde\psi_a^\dagger\gamma_0(\partial_\tau+\mu)\tilde\psi_a
			+{\tilde\psi_a^\dagger\gamma^i\partial_i\tilde\psi_a}
			+\zeta\tilde\phi_a^*\tilde\phi_a+ \sigma\tilde\psi_a^\dagger\tilde\psi_a
			+\frac{N}{8\lambda}\sigma^3-N\frac{\zeta\sigma}{8\lambda}\Big)\Big].
		\end{align}
		Only the zero modes of $\zeta$ and $\sigma$ contribute to the partition function in the leading order in large $N$. Thus ignoring the fluctuations around the zero mode which contribute to subleading order in large $N$, the path integral in the auxiliary fields $\zeta $ and $\sigma$ reduces to ordinary integrals along the variables $\zeta_0$ and $\sigma_0$ which are zero modes of $\zeta$ and $\sigma$ respectively.
		\begin{align}\label{2 integrals}
			\log Z&=\int_{-\infty } ^\infty \int_{-\infty }^\infty d\zeta_0 d\sigma_0 e^{-N\beta V(\frac{\sigma_0^3}{8\lambda}-\frac{\zeta_0\sigma_0}{8\lambda}-\frac{1}{\beta}\log Z_b(\sqrt{\zeta_0},\mu)-\frac{1}{\beta}\log Z_f(\sigma_0,\mu))},
		\end{align}
		where
		\begin{align}\label{Z_b}
			\log Z_b(m_b,-i\hat\mu)&=-\sum_{n=-\infty}^\infty \int \frac{d^2p}{(2\pi)^2}\log[\big(\frac{2\pi n}{\beta}-\hat\mu\big)^2+p^2+m_b^2],\nonumber\\
			&=\frac{1}{6 \pi  \beta ^2} \Big(\beta ^3 m_b^3+3 \beta  m_b (\text{Li}_2(e^{-\beta  (m_b+i\hat\mu )})+\text{Li}_2(e^{-\beta  (m_b- i\hat\mu )}))\nonumber\\
			&\qquad+3 \text{Li}_3(e^{-\beta  (m_b+ i\hat\mu )})+3 \text{Li}_3(e^{-\beta  (m_b- i\hat\mu )})\Big),
		\end{align} 
		and,
		\begin{align}\label{Z_f}
			\log Z_f(m_f,-i\hat\mu)&=\sum_{n=-\infty}^\infty \int \frac{d^2p}{(2\pi)^2}\log[\big(\frac{(2n+1)\pi }{\beta}-\hat\mu\big)^2+p^2+m_f^2]\nonumber,\\
			&=-\frac{1}{6 \pi  \beta ^2}\Big(\beta ^3 m_f^3+3 \beta  m_f (\text{Li}_2(-e^{-\beta  (m_f+ i\hat\mu )})+\text{Li}_2(-e^{-\beta  (m_f-i\hat \mu )}))\nonumber\\
			&\qquad+3 \text{Li}_3(-e^{-\beta  (m_f+i\hat \mu )})+3 \text{Li}_3(-e^{-\beta  (m_f- i\hat\mu )})\Big).
		\end{align}
		The evaluation of  \eqref{Z_b}
		and \eqref{Z_f} involve regularisation of divergent integrals as discussed in the appendix \ref{A}. Note that these integrals are evaluated for imaginary chemical potential but here we will use the analytic continuation by taking $\hat\mu=i\mu$.  \\
		
		Now the integrals over the zero modes of the auxiliary fields in \eqref{2 integrals} are evaluated by the saddle point approximation in $\zeta_0$ and $\sigma_0$. If the saddle point of the integrals are found to be at $\zeta_0=m_b^2$ and $\sigma_0=m_f$,  we have the following pair of saddle point conditions for the integral \eqref{2 integrals} 
		\begin{align}\label{saddle point cond}
			\partial_{\zeta_0} \log Z(\zeta_0,\sigma_0)|_{\zeta_0=m_b^2,\sigma_0=m_f}=0, \qquad \text{and} \qquad \partial_{\sigma_0} \log Z(\zeta_0,\sigma_0)|_{\zeta_0=m_b^2,\sigma_0=m_f}=0.
		\end{align}
		
		These saddle point conditions in $\zeta_0$ and $\sigma_0$ are called the gap equation and further simplified to give the following equations, with the choice $\beta=1$,
		\begin{align}\label{gap eq}
			\frac{m_f}{4\lambda}&=-\frac{m_b}{2\pi}-\frac{1}{2\pi}[\log(1-e^{-m_b+\mu})+\log(1-e^{-m_b-\mu})],\nonumber\\
			\frac{m_{b}^2}{8 \lambda}&=\frac{3 m_f^2}{8\lambda}+\frac{m_f^2}{2\pi}+\frac{m_f}{2\pi}[\log(1+e^{-m_f+\mu})+\log(1+e^{-m_f-\mu})].
		\end{align}
		\\
		
\begin{figure}[t]
			
			\begin{subfigure}{.475\linewidth}
				\includegraphics[width=1\linewidth]{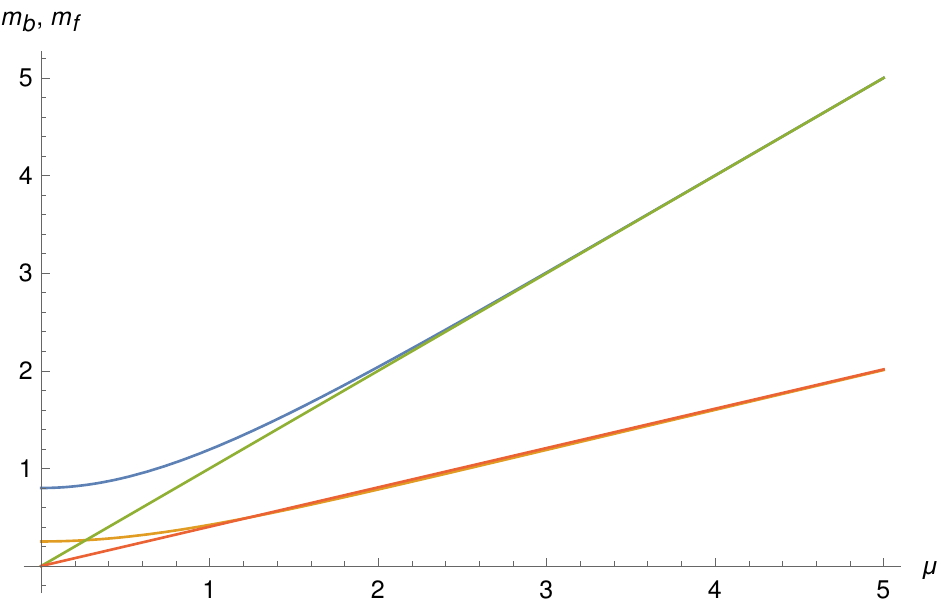}
				\caption{$\lambda=1$}
			\end{subfigure}\hfill 
			\begin{subfigure}{.475\linewidth}
				\includegraphics[width=1\linewidth]{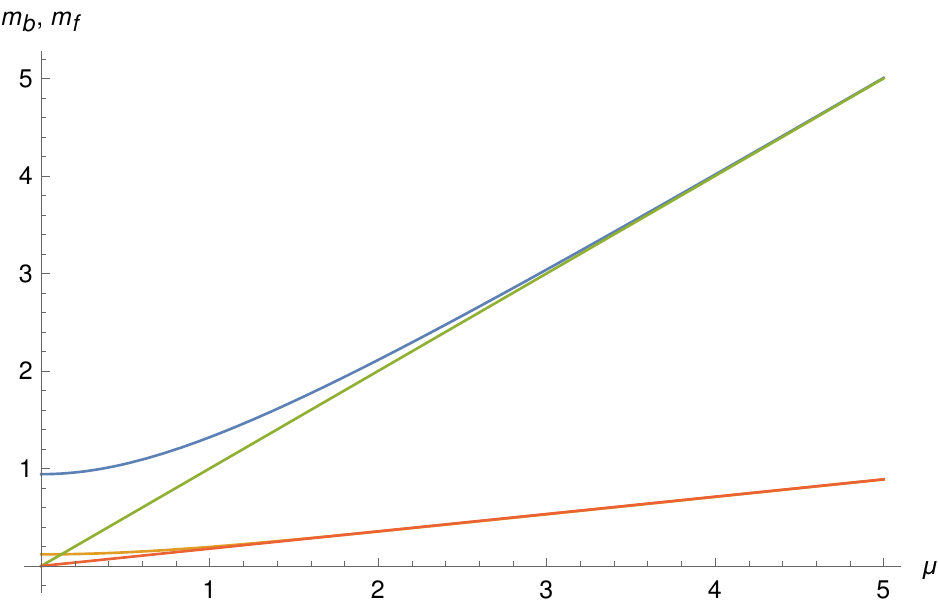}
				\caption{$\lambda=4$}
			\end{subfigure}
			\par\bigskip
			\par\bigskip
			\par\bigskip
			\begin{subfigure}{.475\linewidth}
				\includegraphics[width=1\linewidth]{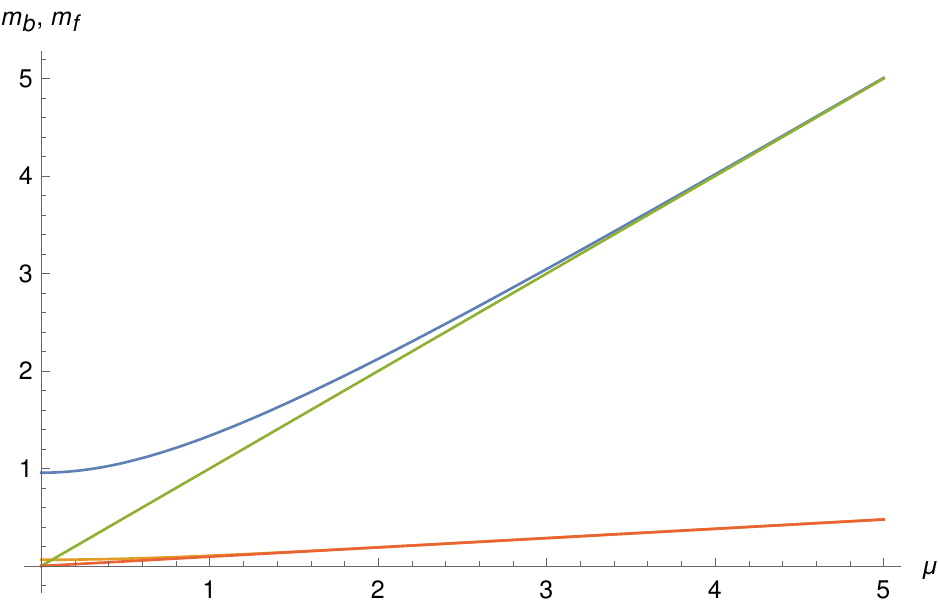}
				\caption{$\lambda=8$}
			\end{subfigure}\hfill 
			\begin{subfigure}{.475\linewidth}
				\includegraphics[width=1\linewidth]{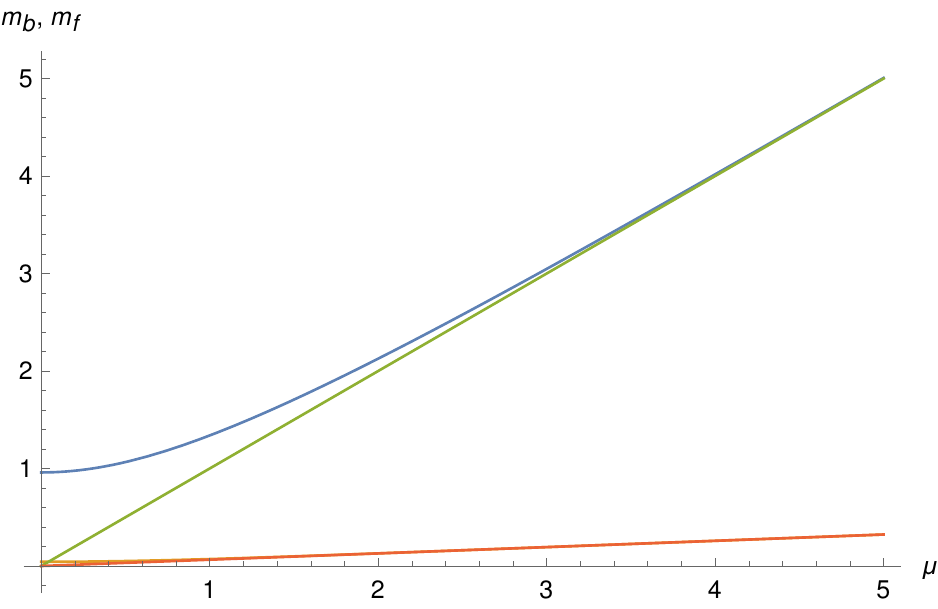}
				\caption{$\lambda$=12}
			\end{subfigure}
			\caption{Large $\mu$ asymptotic solutions for the thermal masses given in \eqref{m_b} and \eqref{m_f} are tested against the numerical solution of the gap equations for different values of the coupling constant $\lambda$. The blue and orange curves refer to the numerical solutions for $m_b$ and $m_f$ of the coupled equations \eqref{gap eq}. While the green and red straight lines represent the asymptotic formulae for $m_b$ and $m_f$  respectively at large $\mu$.}
			\label{fig 1}
		\end{figure}

		The above pair of coupled transcendental equations are studied  numerically at $\mu=0$ in \cite{DeWolfe:2019etx}. But, at large $\mu$ these equations become analytically tractable  and solved in the following manner.  Let us first investigate the asymptotic form of the  1st equation from \eqref{gap eq} at large $\mu$ as given below, note that as we are looking for real and positive solutions of both $m_b$ and $m_f$,
		and we need $0<m_b-\mu<1$ to have consistent solution for $m_b$,
		\begin{align}\label{log mb-mu}
			\frac{\pi  m_f}{2 \lambda }+\log (m_b-\mu )+m_b=0.
		\end{align}
		Now we can solve for $m_b$ from the above equation, and it is given below
		\begin{align}\label{m_b}
			m_b&=\mu +W\left(e^{-\mu -\frac{\pi m_f}{2 \lambda }}\right)
			=\mu+e^{-\mu-\frac{\pi m_f}{2\lambda}}+\cdots,
		\end{align}
		where $W(x)$ is product log function. Now we substitute this solution at the leading order in $\mu$ i.e, $m_b=\mu$ in the 2nd equation of \eqref{gap eq} to obtain
		\begin{align}\label{mu-mf}
			-\frac{\mu ^2}{8\lambda}+\frac{3 m_f^2}{8\lambda}+\frac{m_f^2}{2 \pi }+\frac{m_f (\mu -m_f)}{2 \pi }=0.
		\end{align}
		Thus solving this we obtain the large $\mu$ asymptotic formula for the thermal mass of  fermions to be
		\begin{align}\label{m_f}
			m_f
			=\frac{(\sqrt{4 \lambda ^2+3 \pi ^2}-2 \lambda ) \mu }{3 \pi }+O(e^{-\mu}),
		\end{align}
		keeping only the leading order contribution at large $\mu$. Note that in taking $\mu\to \infty$ limit of  \eqref{mu-mf} we allowed only $m_f<\mu$ as there is no consistent solution otherwise.\\
		The above solutions are valid for $\lambda> 0.$ 
		From the gap equations it is easy to see that at large $\lambda$, the fermion mass vanishes and the model reduces to a combination of $N$ free fermions and $N$ complex scalars at the non-trivial fixed point or the critical point \cite{Filothodoros:2016txa}\footnote{In \cite{Filothodoros:2016txa}, finite imaginary potential was considered which can be analytically continued to real chemical potential.} at finite chemical potential with no fermion boson interaction. These large $\mu$ asymptotic solutions are tested against the numerical solutions of the gap equations and shown in the figure \ref{fig 1}.
		\\

		\section{OPE inversion formula}\label{4}
		In this section we apply the OPE inversion formula on both scalar and fermion two point function of the Wess-Zumino model at the non-trivial fixed point of the theory characterised by the gap equations \eqref{gap eq}. Applying the OPE inversion formula we will evaluate the thermal one point functions of all the different classes of the operators constructed of scalar and fermion bilinears. Note that we will not use the large $\mu$ solutions in this section. We will use the imaginary chemical potential $i\hat \mu$ in the analysis of the OPE inversion formula. By  writing down the thermal two point of scalars and fermions in  the theory at large $N$ in terms of the thermal masses $m_b$ and $m_f $ satisfying the gap equations \eqref{gap eq} and using the OPE inversion formula we will evaluate the thermal one point functions expressed in terms of these thermal masses $m_b$ and $m_f$. We will study the asymptotic limit for these thermal one point functions at the large $\mu$  in the next section, after analytically continuing these answers for the thermal one point functions to real chemical potential. The study being presented in this section is very similar to the study of inversion formula on the two point functions of the conformal fixed point of $U(N)$ scalars and Gross-Neveu model with chemical potential  for the bosonic and fermionic two point functions respectively found in our earlier study for any odd space-time dimension \cite{David:2024naf}. The only difference here is that we have started with more general structure of the OPEs for  the scalar and fermion two point functions given in \eqref{OPE boson} and \eqref{OPE fer}. We have allowed the bosonic bilinear operators to arise in the OPE of the fermionic two point functions and vice versa.
		But as a consequence of the study in this section we will see that the two point function for scalar  only admits the scalar bilinears in the OPE at the large $N$ fixed point and so is the fermion two point function allowing only fermionic bilinears in its OPE.
		\subsection{Scalar two point function}
		Scalar two point function of the twisted field on $S^1\times R^2$ at leading order in large $N$ at imaginary chemical potential $\mu=i\hat\mu$ is given as the following,
		\begin{align}\label{2pt fn WZ}
			g^b(\tau,\vec x)=	\langle \phi^{*} (\tau,\vec{x})\phi(0,0)\rangle=\sum_{n=-\infty}^\infty \int \frac{d^2k}{(2\pi)^2}
			\frac{e^{i(\tilde \omega_n \tau+\vec{k}\cdot\vec{x})}}{\tilde\omega_n^2+\vec{k}^2+m_{b}^2}.
		\end{align}
		where $\tilde\omega_n=\frac{2\pi n}{\beta}-\hat\mu$. Note that the above correlator is computed from the action \eqref{WZ action} using the large $N$ tricks as explained earlier but large $\mu$ limit is not being used. In fact we will  not be using large $\mu$ limit in the two point function. Once we obtain the thermal one point function of the higher spin currents by applying the OPE inversion formula on the two point function, we will directly go for large $\mu$ expansions after analytically continuing to real chemical potential by taking $\hat\mu=-i\mu$ for thermal one point functions as mentioned earlier. \\
		We recast the above expression as the following with the choice $\beta=1$ 
		\begin{align}\label{intermid gb}
			g^b(\tau,\vec x)=\sum_{n=-\infty} ^\infty e^{-i\tau \hat \mu} \int \frac{d\omega d^2k}{(2\pi)^2} \frac{e^{i(\omega\tau+\vec k\cdot \vec x)}}{(\omega-\hat \mu)^2+\vec k^2+m_{b }^2} \delta(\omega-2\pi n).
		\end{align}
		Now by substituting the Fourier series of the Dirac comb as given below
		\begin{align}\label{Dirac comb}
			\sum_{n\in {\mathbb Z} }\delta(\omega-2\pi n)= \frac{1}{2\pi} \sum_{m\in \mathbb{Z}} e^{im\omega},
		\end{align}
		in the \eqref{intermid gb}, we can convert the sum over Matsubara frequencies to the sum over the images in the Euclidean time direction, this method is known as Poisson resummation
		\begin{align}\label{gb image sum}
			g^b(\tau,\vec x)&=\sum_{n=-\infty}^\infty e^{in\hat\mu } \int\frac{d\omega d^2k}{(2\pi)^3} \frac{e^{i\omega(\tau+n)+i\vec k \cdot \vec x}}{\omega^2+\vec k^2+m_{b}^2}\nonumber,\\
			&=\sum_{n=-\infty}^\infty e^{in\hat\mu }  \frac{e^{-m_b\sqrt{(\tau+n)^2+\vec x^2}}}{4 \pi  \sqrt{(\tau+n)^2+\vec x^2}}.
		\end{align}
		
		Now the two point function of the form \eqref{2pt fn WZ} has been studied in \cite{David:2024naf}, to evaluate the thermal one point functions for the model of $U(N)$ scalars at large $N$ with quartic interaction at finite chemical potential using OPE inversion formula  at finite temperature.
		OPE inversion formula is formulated based on the definition of a spectral function $\hat a(\Delta,l)$ where $l$ is the spin of the operators which occur in the OPE of the two point function. The spectral function $\hat a(\Delta,l)$ admits poles at the dimension of the operators($ \Delta_\mathcal{O} $) or the operator spectrum in the OPE at a certain value of spin($l$). From the negative residue at these poles at $\Delta=\Delta_\mathcal{O}$ the thermal one point functions of spin-$l$ operators of dimension $\Delta_\mathcal{O}$ are being read out  as shown below,

		\begin{align} 
			a_\mathcal{O}[n,l]&=-\hat a( \Delta_{} , l)|_{{\rm Res\ at\ }\Delta=\Delta_\mathcal{O}=2n+l+2\Delta_\phi} .
		\end{align}
		Now, we define,
		\begin{align}
			z=\tau+i x=rw \qquad {\rm and} \qquad \bar z=\tau-ix=rw^{-1},
		\end{align}
		where $\tau$ is the imaginary time direction as usual and $x$ is the spatial distance in two dimensional plane. 
		And in the complex $w$-plane the spectral function has contribution due to the branch cut discontinuity of the two point function in the $w$ plane denoted by $\hat a_{\rm disc} ( \Delta_{}, l)$ and the circular arc at infinity $\hat a_{\rm arc} ( \Delta_{}, l)$ as given below,
		\begin{align}
			\hat a(\Delta,l)= - \big(\hat a_{\rm disc} ( \Delta_{}, l) + \theta( l_0 -l) \hat a_{\rm arc} ( \Delta_{} ,l)\big),
		\end{align}
		The arc contribution vanishes unless the angular momenta is bounded above, up to a given angular momentum $l_0$. Here 
		$ \hat a_{\rm disc}(\Delta,l) $ is given by,
		\begin{align}\label{a disc}
			&\hat a_{\rm disc}(\Delta,l)=K_l \int_0^1 \frac{d\bar z}{\bar z}\int_1^{1/\bar z}\frac{dz}{z} (z \bar z)^{\Delta_\phi-\frac{\Delta}{2}-\nu}(z-\bar z)^{2\nu} F_l\bigg(\sqrt{\frac{z}{\bar z}}\bigg) {\rm Disc} [{g^b}(z,\bar z)], \\
			&\text{where,}\quad \nonumber
			K_l=\frac{\Gamma(l+1)\Gamma(\nu)}{4\pi \Gamma(l+\nu)},
			\qquad F_l(w) = w^{l+ 2\nu } {}_2 F_1( l + 2\nu , \nu , l + \nu +1, w^2) .
		\end{align}
		and,
		\begin{align} \label{Disc}
			{\rm Disc}[ g^b(z, \bar z) ] = \frac{1}{i} \big(  g^b( z +i \epsilon , \bar z) - g^b( z-i\epsilon, \bar z) \big) .
		\end{align}
		While the rest of the contribution $ a_{\rm arc}(\Delta,l) $ involves the contribution at infinity in the $z$-plane. 
		
		The 
		integration along a circle of infinite radius in the complex $ w $ plane is given by 
		\begin{align}\label{arc part}
			\hat a_{\rm arcs} (\Delta_{},  l)  &=  2 K_l   \int_0^1 \frac{dr}{r^{\Delta_{}  +1 - 2\Delta_\phi} }  \times   \oint \frac{dw}{i w} \lim_{|w| \rightarrow \infty} 
			\left[ \Big( \frac{ w - w^{-1} }{i} \Big)^{2\nu} 
			F_l(w^{-1}) e^{i\pi\nu}  {g}^b( r, w) 
			\right].
		\end{align}
		Now from the propagator \eqref{gb image sum} and \eqref{Disc} we have
		\begin{align}
			{\rm Disc} [g^b(z,\bar z)]=\frac{1}{2 \pi}\sum_{\substack{m=-\infty\\m\ne 0}}^\infty e^{-i m\hat \mu}\,\frac{\cos \left(m_{th} \sqrt{(z-m) (m-\bar z)}\right)}{\sqrt{(z-m) (m-\bar z)}}.
		\end{align}
		We substitute this expression into the  equation \eqref{a disc} to obtain, 
		\begin{align}\label{a++a-}
			\hat a_{\rm disc}(\Delta,l)=\hat a_{\rm disc}^{+}(\Delta,l)+\hat a_{\rm disc}^{-}(\Delta,l).
		\end{align}
		Where $a^{\pm}_{\rm disc}(\Delta,l)$ is defined so that the contribution due to $m>0$ and $m<0$ keep separated, as given below
		\begin{align}
			\hat a_{\rm disc}^{\pm}(\Delta,l)=\frac{K_l}{2\pi}\sum_{m=1}^{ \infty}\int_0^1 \frac{d\bar z}{\bar z}\int_m^{{\rm max}(m,\frac{1}{\bar z})}\frac{d z}{ z}\frac{ (z-\bar z) F_l\left(\sqrt{\frac{\bar z}{z}}\right) (z \bar z)^{-\frac{\Delta }{2}} e^{\mp i \hat \mu  m} }{ \sqrt{(z\mp m) (\pm m-\bar z)}}\nonumber\\
			\times\cos \left(m_{\rm th} \sqrt{(z\mp m) (\pm m-\bar z)}\right).
		\end{align}
		It is easy to see that,
		\begin{align}\label{a-}
			a_{\rm disc}^{-}(\Delta,l)=(-1)^la^+_{\rm disc}(\Delta,l)|_{\hat \mu\to -\hat \mu}.
		\end{align}
		Now using the transformation, 
		\begin{align}
			\bar z=z'\bar z'\qquad {\rm and} \qquad z=mz',
		\end{align}
		We obtain, 
		\begin{align}
			\hat a^+_{\rm disc}(\Delta,l)=-\frac{K_l}{2\pi}\sum_{m=1}^\infty\int_0^1 d\bar z\int_1^{{\rm max}(1,\frac{1}{m\sqrt{\bar z}})} dz\frac{(\bar z-1) e^{-i \mu  m} F_l(\sqrt{\bar z})  \cos \left(m m_{\rm th} \sqrt{(1-z) (z \bar z-1)}\right)}{\bar z \left(m^2 z^2 \bar z\right)^{\frac{\Delta }{2}} \sqrt{(1-z) (z \bar z-1)}}.
		\end{align}
		Now we expand the it at  small $\bar z$ and keep only the leading order contribution, performing the above integral from $1$ to $\infty$ in $ z$ we get
		\begin{align} \label{hat a}
			\hat	a_{\rm disc}^{+(0)}(\Delta,l)&=\frac{2K_l}{\pi}\sum_{m=1}^\infty\frac{e^{-im\hat  \mu}}{m^\Delta(-\Delta +l+1)}\int_0^\infty dy \frac{   \cos (m m_{\rm th} y)}{(1+y^2)^\Delta},\nonumber\\
			&=\frac{2K_l}{\pi}\sum_{m=1}^\infty\frac{e^{-im\hat  \mu}}{\sqrt{m}} 
			\frac{\sqrt{\pi } 2^{\frac{1}{2}-\Delta } m_b^{\Delta -\frac{1}{2}} K_{\Delta -\frac{1}{2}}(m m_b)}{(-\Delta +l+1)\Gamma (\Delta )},\\
			\nonumber{\rm for}\qquad l>0.
		\end{align}
		It is clear from the above expression that the spectral function admits pole at $\Delta=l+1$ which is the dimension of the bosonic operators  ${\cal O}[0,l]$ and ${\cal \tilde O}_+[0,l]$ as these two operators were combined in the 2nd term of the OPE \eqref{ph ph OPE}.  And the sub-leading order by order corrections in $\bar z$ will give the poles arising due to higher twist($n>0$) operators from the combination of the same classes ${\cal O}[n,l]$  and ${\cal \tilde O}_+[n,l]$ which can be verified straightforwardedly. This pole structure discards the possibility of the presence of the class of ferminic operator ${\cal \tilde  O}_0[n,l]$ in the bosonic OPE for this model at large $N.$ Thus, using this pole structure in \eqref{hat a} and the equation \eqref{ph ph OPE} we get
		\begin{align}
			a^b_{{\cal \tilde O}_0}[0,l]=0,
		\end{align} 
		which implies that
		\begin{align}
			C_{\phi\phi{\cal \tilde O}_0[0,l] }=0, \qquad {\rm at \ large \ }N,
		\end{align}
		As the thermal one point functions of the fermionic operator ${\cal \tilde O}_0[0,l]$ is non-vanishing.  We will evaluate thermal one point function of fermionic operators in the next subsection \ref{OPE inver}, by examining the OPEs for the fermion two point functions. The OPE coefficients  $C_{\phi\phi{\cal \tilde O}_+[0,l]}$ and $C_{\phi\phi{\cal \tilde O}_-[0,l]}$ can be non-zero in principle. But we know that $C_{\phi\phi{\cal \tilde O}_+[0,l]}=C_{\phi\phi {\cal \tilde O}_-[0,l]}=0 $  at the Gaussian fixed point where $\lambda=0$ or $m_b=m_f=0$ as well as at the non-trivial fixed point at strong coupling $\lambda\to \infty$. Thus we  expect that it is true for all $\lambda$. But we will explicitly show that this is indeed true for $l=1$ and $l=2$ in the section \ref{sec 6} for all $\lambda$. We evaluate the spin-1  and  spin-2 current from the partition function of the theory and comparing these against the thermal expectations $a_{\cal O}[0,1] $ and $a_{\cal O}[0,2]$ respectively we can evaluate $C_{\phi\phi{\cal O}[0,1]},C_{\phi\phi{\cal O}[0,2]}$ for arbitrary coupling. We will observe that $C_{\phi\phi{\cal O}[0,1]},C_{\phi\phi{\cal O}[0,2]}$ do not depend on $\lambda$ at the leading order in large $N$ which implies that  $C_{\phi\phi{\cal \tilde O}_+[0,1]}=C_{\phi\phi {\cal \tilde O}_+[0,2]}=0  $ as explained in section \ref{sec 6}. We extrapolate this observation for any $l$, i.e., $a^b_{{\cal \tilde O}_+}[0,l]=0$. Finally taking the negative residue of the spectral function obtained at \eqref{hat a} and using \eqref{a++a-} we evaluate the thermal one point function of the operator ${\cal O}[0,l]$ to be 
		\begin{align}\label{a O}
			a^b_\mathcal{O}[0,l]&=-(a^{+{(0)}}_{\rm disc}+a^{-(0)}_{\rm disc})|_{{\rm Res\ at\ }\Delta=l+1}\nonumber\\
			&=\frac{K_l}{ l!}\sum_{n=0}^l\frac{ m_b^{l-n} (l-n+1)_{2 n} }{2^{l+n} n! }\times\Big(\text{Li}_{n+1}(e^{-m_b-i \hat \mu })+(-1)^l \text{Li}_{n+1}(e^{-m_b+i \hat \mu })\Big) ,\\
			{\rm for }\  l>0.\nonumber
		\end{align}

		\subsection{Fermion two point functions}\label{OPE inver}
		We consider the two point correlator of the twisted fermion fields as given below,
		\begin{align} \label{2 pt fn fer}
			\langle \psi_\alpha^\dagger(x) \psi_\beta(0)\rangle =\frac{i}{2}\sum_{k_0\in (2n+1) \pi }\int \frac{d^{2}k}{(2\pi)^{2 }}\frac{\gamma^\mu_{\alpha\beta} \tilde k_\mu+im_{ f}\delta_{\alpha\beta}}{\tilde k^2+m_{ f}^2} e^{i\tilde k. x}.
		\end{align}
		with $\tilde k_\nu=k_\nu-\hat \mu\delta_{\nu 0} $ and $n\in \mathbb{Z}$. $ \alpha,\beta $ denoting the spinor indices. 
		Recall that $m_f$ is the  thermal mass of the fermions satisfying the gap equations \eqref{gap eq}. Note that  in this propagator we keep the chemical potential to be imaginary as given by $\mu=i\hat \mu.$\\
		
		Now we will be studying this fermionic correlator in three different channels as was defined in \eqref{diff ch fer} and apply the OPE inversion formula on each of these channels  to evaluate the thermal one point functions of three different classes of operators constructed out of fermion bilinears.
		\subsubsection*{$g_1(x)$}The expectation value $a_{{\cal O}_0} [0, l]$     is contained in the OPE  
		of the correlation function $g_1(x)$. Now using \eqref{2 pt fn fer} and \eqref{diff ch fer} we can evaluate $g_1{(x)}$ very easily and we get it to be the following
		\begin{align}
			g_1(x)=\langle \psi_\alpha^\dagger(x)\psi_\alpha(0)\rangle=-m_f \sum_{\omega_m\in (2m+1)\pi}e^{-i\hat \mu\tau}\int\frac{d^2k}{(2\pi)^2} \frac{e^{i\omega_m \tau+i\vec k\cdot \vec x}}{(\omega_m-\hat \mu)^2+\vec{k}^2+m_f^2},
		\end{align}
		for $m\in \mathbb{Z}$.\\
		
		Now to express $g_1(x) $ completely in the position space, we recast the above expression as follows
		\begin{align}
			g_1(x)=-m_f \sum_{\omega_m\in {(2m+1)\pi}} e^{-i\hat\mu\tau} \int d\omega \delta(\omega-\omega_m) \int\frac{d^2k}{(2\pi)^2} \frac{e^{i\omega\tau} e^{i\vec k \cdot \vec x}}{(\omega-\hat \mu)^2+\vec{k}^2+m_f^2},
		\end{align}
		With the use of the Fourier series of the Dirac comb as
		\begin{align}
			\sum_{\omega_m\in (2m+1)\pi }\delta(\omega-\omega_m)= \frac{1}{2\pi} \sum_{m\in \mathbb{Z}} e^{im\omega}(-1)^m,
		\end{align}
		and using the change of variable $\omega'=\omega-\hat \mu$ in the next step, we obtain,
		\begin{align}
			g_1(x)=-m_f\sum_{m\in \mathbb{Z}}\frac{(-1)^m e^{im\hat \mu}}{(2\pi)^3} \int d\omega d^2k \frac{e^{i\omega (\tau+m)}e^{i\vec{k}\cdot\vec x}}{\omega^2+\vec k^2+m_f^2}.
		\end{align}
		Now using the standard method of Poisson resummation we can rewrite the sum over the matsubara frequencies in the above expression as the sum over images as follows
		\begin{align}\label{g_1}
			g_1(z,\bar z)=-m_f \sum_{m\in \mathbb{Z}}(-1)^me^{-i\hat\mu m}  \frac{e^{-m_f \sqrt{(m-z)(m-\bar z)}}}{4\pi \sqrt{(m-z)(m-\bar z)}},
		\end{align}
		Applying the Euclidean inversion formula \eqref{a disc} on $g_1(z,\bar z)$ given in \eqref{g_1}  in the similar manner as shown for the bosonic correlator  in the previous subsection we obtain the following the spectral function for the one point function $\hat a^{+(0)}(\Delta,l)$. 
		\begin{align}\label{spectr fer g_1}
			\hat a^{+(0)}(\Delta,l)=K_lm_f\sum_{m=1}^\infty\frac{(-1)^{m+1}  z^{\frac{l+1}{2}} e^{-i\hat \mu  m} \left(m^2 z^3\right)^{\frac{1}{2}-\frac{\Delta }{2}} \cos \left(m m_f \sqrt{z-1}\right)}{\pi  \sqrt{z-1} (-\Delta +l+2)}.
		\end{align}
		Taking the negative residue of the above expression for $\hat a^{+(0)}(\Delta,l)$ we get
		\begin{align}\label{a O 0 disc}
			-\hat a^{+(0)}[0,l]|_{\text{Res at }\Delta=l+2}=K_l\sum_{m=1}^\infty\frac{2^{\frac{1}{2}-l} (-1)^{m+1} m_f^{l+\frac{3}{2}} e^{-i\hat \mu  m} K_{l+\frac{1}{2}}(m m_f)}{\sqrt{\pi m } \Gamma (l+1)}.
		\end{align}
		Finally using finite sum representation of $K_{l+\frac{1}{2}}(x)$ for $l\in \mathbb Z$, we can perform the  infinite sum over $m$ to obtain the thermal one point function to be
		\begin{align}\label{a O 0 f}
			a^f_{{\cal\tilde O}_0}[0,l]	&=-K_l\sum_{n=0}^l \frac{ m_f^{l-n+1} (l-n+1)_{2 n} \big(\text{Li}_{n+1}\left(-e^{-i \mu -m_f}\right)+(-1)^l \text{Li}_{n+1}\left(-e^{i \mu -m_f}\right)\big)}{2^{l+n}n! \Gamma (l+1)},\nonumber\\
			&{\rm for \ }l>0.
		\end{align}
		Note that the pole structure of $\hat a^{+(0)}(\Delta,l)$ in \eqref{spectr fer g_1} clearly rules out the possibility of the  presence of the bosonic operator ${\cal O}[0,l]$ in the OPE  \eqref{g_1 ope} as $\hat a^{+(0)}(\Delta,l)$ has pole only at $\Delta=l+2$ which stands for the fermionic bilinear operators of the kind ${\cal O}_0[0,l]$. Thus from this pole structure in \eqref{a O 0 f} and the OPE \eqref{g_1 ope} we can conclude that
		\begin{align}\label{C ppO=0}
			a_{\cal O}^f[0,l]=0.
		\end{align}
		And now using \eqref{structure const fer} and using the fact that the thermal one point function of the scalar bilinears denoted by $b_{\cal O}$ is non-zero as we have seen in equation \eqref{a O}, we have
		\begin{align}
			C_{\psi\psi{\cal O}[0,l]}=0,
		\end{align}
		at large $N$.
		\subsection*{$g_2(x)$}
		From \eqref{2 pt fn fer} and \eqref{diff ch fer} we can evaluate the correlator $g_2(x) $ as the following
		\begin{align}\nonumber \label{fer 2 pt g_2}
			g_2(\tau,\vec{x})&=\frac{i}{|x|}\sum_{k_0\in\pi(2n+1)}\int \frac{d^{2}k}{(2\pi)^{2 }}\frac{\tilde k_\mu x^\mu}{\tilde k^2+m_{th}^2} e^{i\tilde k x}, \\
			&=\frac{1}{|x|}x^\mu\partial_\mu\sum_{k_0\in\pi(2n+1)}\int \frac{d^{2}k}{(2\pi)^{2 }}\frac{e^{i\tilde k x}}{\tilde k^2+m_{th}^2} .
		\end{align}
		The  method of  Poisson re-summation formula as was used for $g_1(z,\bar z)$ transforms the above  Matsubara sum into a sum over images to give
		\begin{align} \label{g_2 inv}
			g_2(z, \bar z) 
			=\sum_{m\in \mathbb{Z}}&(-1)^{m+1}e^{-im \hat \mu}
			\bigg[-\frac{m}{2} \sqrt{\frac{z}{\bar z}} - \frac{m}{2} \sqrt{\frac{\bar z}{ z}}
			+\sqrt{z \bar z}\bigg] \times\nonumber\\
			&
			\frac{e^{-m_f \sqrt{(m-z)(m-\bar z)}} (m_f \sqrt{(m-z)(m-\bar z)}+1)}{4 \pi  [(m-z)(m-\bar z)]^{3/2}}. 
		\end{align}
		Using the method of OPE inversion formula on the correlator $g_2(z,\bar z)$ given above, we can obtain the thermal one point functions of arbitrary spin fermion bilinears. The method has already been implemented in \cite{David:2023uya} and the result was obtained at the equation  (4.26) of \cite{David:2023uya} for any odd dimension(we put $k=1$ in the result of \cite{David:2023uya} as the notation used there is $d=2k+1$). Note that the two point function $g_2(z,\bar z)$ has an extra $e^{-im\hat\mu}$ compared the correlator studied in equation (4.13) of \cite{David:2023uya}, thus taking the factor of $e^{-im\hat\mu}$ into account to this calculation we have
			
			\begin{align}\label{a O+}
				a^f_{\mathcal{\tilde O}_+}[n=0,l]&=\frac{l}{{ \pi \big(\frac{1}{2}\big)_l}}{ \,\sum _{n=0}^l \frac{m_f^{l-n} (l-n+1)_{2 n}}{2^{n+l+2} n!}} \left(\text{Li}_{n+1}(-e^{-i\hat\mu -m_f})+(-1)^l \text{Li}_{n+1}(-e^{-m_f+i\hat\mu })\right),
			\end{align}
			for $l>0.$\\
			
			We used the fact that $C_{\psi\psi{\cal O}[0,l]}=0$ from \eqref{C ppO=0} i.e., scalar bilinear operators of the kind ${\cal O}[0,l]$ are not present in the OPE of the fermion two point function given in \eqref{g2 gen struct}.\\
			
			Applying the inversion formula on the correlator $g_2(z,\bar z)$ we also have obtained the following result in the equation (4.33) of \cite{David:2023uya} with $d=3$(or $k=1$ as the results obtained there in odd space time dimensions $d=2k+1,\ k\in \mathbb Z$),
			\begin{align}\label{a+ + a- form g2}
				&	 a^f_{{\cal \tilde O}_+}[1, l] + \frac{ 2( l +1)}{ ( l +2) (  l+3 ) } a^f_{{\cal\tilde O}_-}[0, l]\nonumber \\
				&	=    
				\frac{ ( l+2) }{  2\pi ( 3+ 2l ) ( \frac{1}{2} )_l } 
				\left( \frac{m_f}{2} \right)^{l +2 } 
				\sum_{n=0}^{l } \frac{ (l +1 -n)_{2n}}{ (2 m_f) ^n n!} ({\rm Li}_{n+1} ( - e^{-m_f-i\hat\mu} )+(-1)^l {\rm Li}_{n+1} ( - e^{-m_f+i\hat\mu} )),\nonumber\\
				& \qquad\qquad{\rm for}\ \  l>0.
			\end{align}
			
			Note that here we have an extra $e^{-im\hat\mu} $ in \eqref{g_2 inv} compared to the correlator given in the equation (4.13) of \cite{David:2023uya}. Thus we have accounted for the required modifications due to this $e^{-im\hat \mu}$ in the above expression.
			\subsection*{$g_3(x)$}
			Using \eqref{2 pt fn fer} and \eqref{diff ch fer} we can write down $g_3( x) $ as given below
			\begin{align}
				g_3(x)&=\partial^\mu\partial_\mu \sum_{k_0\in (2n+1)\pi} \int \frac{d^2k}{(2 \pi)^2} \frac{e^{i\tilde k \cdot x}}{\tilde k^2+m_f^2}\nonumber,\\
				&=\partial^\mu\partial_\mu \sum_{m\in \mathbb{Z}}(-1)^me^{i\hat\mu m}  \frac{e^{-m_f \sqrt{(m+\tau)^2+{\vec x}^2}}}{4\pi \sqrt{(m+\tau)^2+{\vec x}^2}}.
			\end{align}
			Thus we have
			\begin{align}\label{g_3}
				g_3(x)=m_f^2\sum_{m\in\mathbb{Z}}\frac{(-1)^m  e^{-m_f \sqrt{(m+\tau )^2+{\vec x}^2}+i \hat\mu  m}}{4 \pi  \sqrt{(m+\tau )^2+{\vec x}^2}}.
			\end{align}
			Again by using $C_{\psi\psi \cal O}=0 $ from \eqref{C ppO=0} the entire analysis of the inversion formula on  $g_3(x)$ reduces to the computation of the thermal one point functions of fermion bilinears from the $g_3(x)$ for the large $N$ Gross-Neveu model performed  in \cite{David:2023uya}, except the fact that one has to keep track of the $e^{im\mu }$ in the correlator \eqref{g_3} compared to the the correlator given in equation (4.33) of \cite{David:2023uya} with $d=3$(or $k=1$ as the notation used there $d=2k+1$)\footnote{Note that there was a typo in the equation (4.33) of \cite{David:2023uya}, in the denominator of this formula $(k+\frac{1}{2})_l$ is not correct, it should be replaced by $(k-\frac{1}{2})_l$.}.  And keeping this $e^{im \mu}$ in that calculation we get 
			\begin{align}\label{a+  +a- from g3}
				&2 a^f_{{\cal\tilde O}_+}[1,l]+(l+1)a^f_{{\cal\tilde O}_-}[0,l]\nonumber \\
				&=\frac{1}{\pi (\frac{1}{2})_l}
				\Big(\frac{m_f}{2}\Big)^{l+2} \sum_{n=0}^l\frac{(l-n+1)_{2n}}{(2m_f)^nn!}({\rm Li}_{n+1}(-e^{-m_f-i\hat\mu})+(-1)^l {\rm Li}_{n+1}(-e^{-m_f+i\hat\mu})).\\
				&{\rm for}\qquad l>0.\nonumber
			\end{align}
			We can solve \eqref{a+ + a- form g2} and \eqref{a+  +a- from g3} to obtain $a_{{\cal O}_-}[0,l]$ and $a_{{\cal O}_-}[1,l]$, although we do not show the calculation as it is not so illustrative.\\
			
			\subsection*{Arc Contributions}
			In \cite{David:2023uya,David:2024naf} we have seen that the contribution to the inversion formula due to the circular arc at infinity given in \eqref{arc part} is non-vanishing only for $l=0$ and it vanishes for $l>0$ for all the scalar and fermion correlators. Thus for $l>0$ it is enough to consider disc contribution only.\\
			For $l=0$, it is necessary to evaluate the arc contribution. Here, we demonstrate that the gap equations can be correctly recovered from the thermal one-point functions of the scalar and fermion currents at 
			$l=0$.
			For the scalar bilinear at $l=0$ or the thermal expectation of $\phi^2$ we must include the arc contribution obtained using the formula \eqref{arc part}. For scalar operator ${\cal O}[0,0] $ it has been computed in \cite{David:2024naf} and given by\footnote{See equation (2.50) in \cite{David:2024naf}.}
			\begin{align}\label{arc boson}
				-\hat a_{\rm arcs}(\Delta,0)|_{{\rm Res\ at} \ \Delta=1}=-\frac{m_b}{4\pi}.
			\end{align}
			Now combining the arc contribution \eqref{arc boson} and the disc contribution given in \eqref{hat a} at $l=0$ we have
			\begin{align}\label{a O [0,0]}
				a_{\cal O}[0,0]&\nonumber=-(\hat a_{\rm disc}+\hat a_{\rm arcs})|_{{\rm Res\ at} \ \Delta=l+1},\\&=-\frac{m_b}{4\pi}-\frac{1}{4\pi}(\log (1-e^{-i\hat\mu -m_b})+\log (1-e^{i\hat\mu -m_b})).
			\end{align}
			Now from the partition function \eqref{part fn} at the saddle point described in \eqref{saddle point cond} we have
			\begin{align}
				a_{\cal O}[0,0]=\langle \phi^2\rangle_\beta= \frac{m_f}{8\lambda}.
			\end{align}
			Thus using the above equation and \eqref{a O [0,0]} we recover the correct gap equation as was given in the 1st line of \eqref{gap eq}
			\begin{align}
				\frac{\pi m_f}{2\lambda}+m_b+\log (1-e^{-i\hat\mu -m_b})+\log (1-e^{i\hat\mu -m_b})=0.
			\end{align}
			\\
			
			For the one point function of the fermion bilinear ${\cal O}_0[0,0]$, the arc contribution has also been evaluated in \cite{David:2023uya} and it is given by \footnote{See equation (4.49) in \cite{David:2023uya}, note that the presence of the chemical potential $-i\hat\mu$ doesn't alter this answer, we kept track of an overall -ve sign in our correlator $g_1(z,\bar z)$ in \eqref{g_1} compared to that used in \cite{David:2023uya}.}
			\begin{align}\label{arc fer}
				-\hat a_{\rm arcs}(\Delta,0)|_{{\rm Res\ at} \ \Delta=2}=\frac{m_f^2}{4\pi}.
			\end{align}
			Combining arc contribution \eqref{arc fer} and disc contribution \eqref{a O 0 disc} we obtain for 
			\begin{align}\label{a O 0 [0,0]}
				a_{{\cal O}_0}[0,0]&\nonumber=-(\hat a_{\rm disc}+\hat a_{\rm arcs})|_{{\rm Res\ at} \ \Delta=2},\\&=\frac{m_f^2}{4\pi}+\frac{m_f}{4\pi} \left(\log \left(1+e^{-m_f+i \mu }\right)+\log \left(1+e^{-m_f-i \mu }\right)\right).
			\end{align}
			From the partition function  \eqref{part fn} at the saddle point \eqref{saddle point cond}, we also have
			\begin{align}
				a_{{\cal O}_0}[0,0]=2 \langle \psi^\dagger\psi\rangle_\beta=-\frac{3m_f^2}{8\lambda}+\frac{m_b^2}{8\lambda}.
			\end{align}
			Note that the factor of $2$ in the above equation arises as we had normalised the 2-point function in \eqref{2 pt fn fer} with a factor of $\frac{1}{2}$. Substituting \eqref{a O 0 [0,0]} in the above equation, we recover the correct gap equation as was given in the 2nd line of \eqref{gap eq}.
			\begin{align}
				\frac{m_f^2}{2\pi}+\frac{m_f}{2\pi} (\log \left(1+e^{-m_f+i \mu }\right)+\log (1+e^{-m_f-i \mu }))+\frac{3m_f^2}{8\lambda}-\frac{m_b^2}{8\lambda}=0.
			\end{align}
			This gap equation can similarly be obtained by calculating $a_{{\cal O}_-}[0,0]$, though we do not show it here.
			\section{Thermal one point functions at large $\mu$} 
			\label{sec 5}
			In this section, we use the analytic answers for the thermal masses for complex scalars and fermions at large $\mu$ obtained in \eqref{m_b} and \eqref{m_f} respectively in the expressions for the thermal one point functions evaluated in the section \ref{4} using OPE inversion formula and evaluate the systematic expansions for the thermal one point functions at large $\mu$. 
			\subsection*{Scalar bilinears}
			
			\begin{figure}[hbt!]
				
				\begin{subfigure}{.475\linewidth}
					\includegraphics[width=1\linewidth]{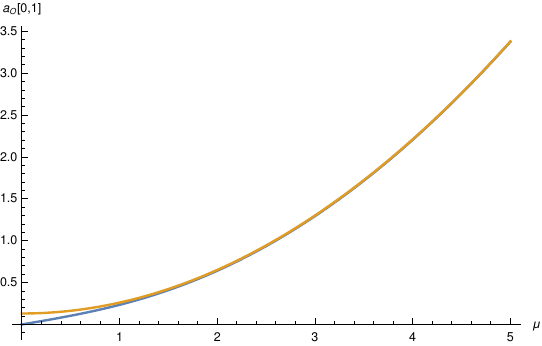}
					\caption{$\lambda=1$}
				\end{subfigure}\hfill 
				\begin{subfigure}{.475\linewidth}
					\includegraphics[width=1\linewidth]{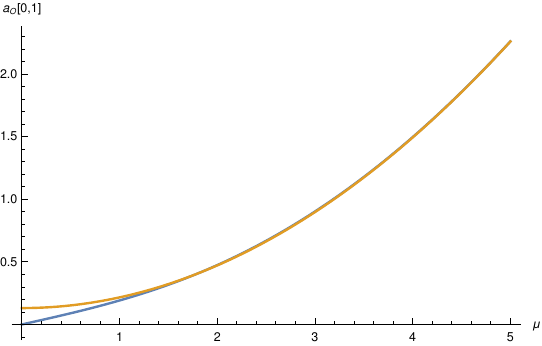}
					\caption{$\lambda=4$}
				\end{subfigure}
				\par\bigskip
				\begin{subfigure}{.475\linewidth}
					\includegraphics[width=1\linewidth]{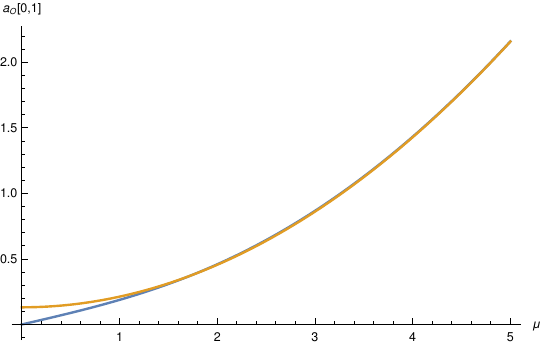}
					\caption{$\lambda=8$}
				\end{subfigure}\hfill 
				\begin{subfigure}{.475\linewidth}
					\includegraphics[width=1\linewidth]{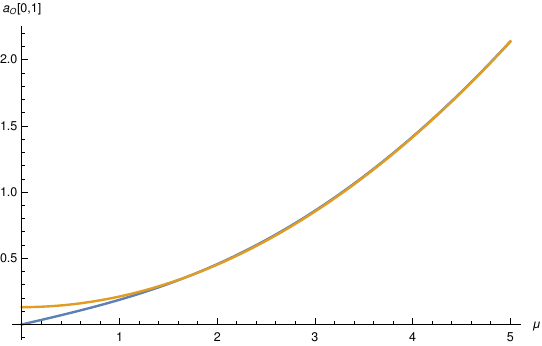}
					\caption{$\lambda=12$}
				\end{subfigure}
				
				\caption{We compare the asymptotic formula \eqref{asymp formula bosons} for $a_{{\cal O}}[0,1]$ against the the numerical value of the $a_{{\cal O}}[0,1]$ obtained from the \eqref{a O bosons} with the use numerical solutions of the gap equation \eqref{gap eq} for different values of the coupling constant $\lambda$. Blue curve denotes the numerical answer while the orange curve stands for the asymptotic formula \eqref{asymp formula bosons}. We kept $\beta=1$ as usual. }
				\label{a O l=1}
			\end{figure}
			\begin{figure}[t]
				
				\begin{subfigure}{.475\linewidth}
					\includegraphics[width=1\linewidth]{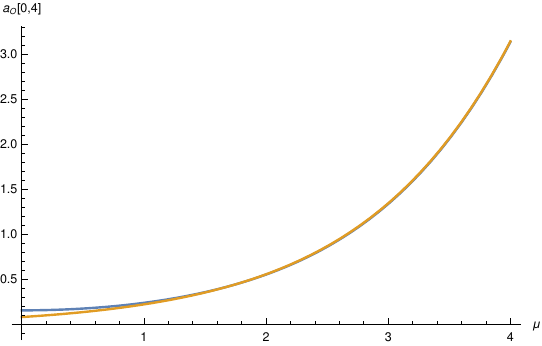}
					\caption{$\lambda=1$}
				\end{subfigure}\hfill 
				\begin{subfigure}{.475\linewidth}
					\includegraphics[width=1\linewidth]{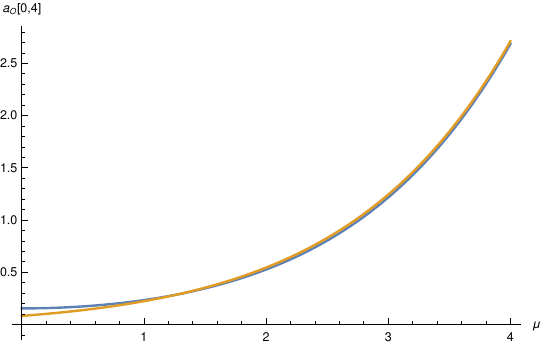}
					\caption{$\lambda=4$}
				\end{subfigure}
				\par\bigskip
				\begin{subfigure}{.475\linewidth}
					\includegraphics[width=1\linewidth]{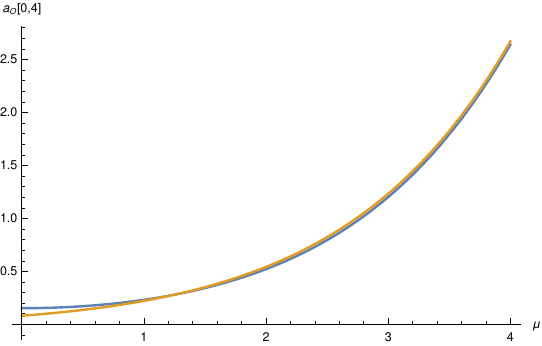}
					\caption{$\lambda=8$}
				\end{subfigure}\hfill 
				\begin{subfigure}{.475\linewidth}
					\includegraphics[width=1\linewidth]{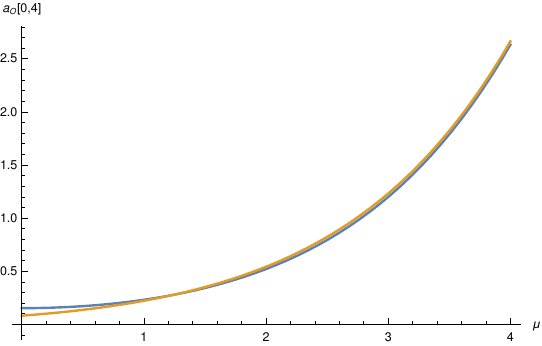}
					\caption{$\lambda=12$}
				\end{subfigure}
				
				\caption{We compare the asymptotic formula \eqref{asymp formula bosons} for $a_{{\cal O}}[0,4]$ against the the numerical value of the $a_{{\cal O}}[0,4]$ obtained from the \eqref{a O bosons} with the use numerical solutions of the gap equation \eqref{gap eq} for different values of the coupling constant $\lambda$. The blue curve denotes the numerical answer while the orange curve stands for the asymptotic formula \eqref{asymp formula bosons}, note that $\beta=1.$  }
				\label{a O l=4}
			\end{figure}
			Let us first focus on the thermal one point functions of the bosonic operators ${\cal O}[0,l]$ evaluated at \eqref{a O}, analytically continuing  back to real  chemical potential by using $\hat\mu=i\mu$, we have
			\begin{align}\label{a O bosons}
				a_\mathcal{O}[0,l]
				&=\frac{K_l}{ l!}\sum_{n=0}^l\frac{ m_b^{l-n} (l-n+1)_{2 n} }{2^{l+n} n! }\times\Big(\text{Li}_{n+1}(e^{-m_b+ \mu })+(-1)^l \text{Li}_{n+1}(e^{-m_b-  \mu })\Big), \nonumber\\
				{\rm for }\  l>0.
			\end{align}
			Now, at large $\mu$ we can write down the asymptotic form of the above expression as presented below
			\begin{align}
				a_{\cal O}[0,l]=\frac{K_l}{l!}\Big[-\frac{m_b^l}{2^l}\log (m_b-\mu)+\sum_{n=1}^\infty \frac{ m_b^{l-n} (l-n+1)_{2 n} }{2^{l+n} n! }\zeta(n+1)+O(e^{-\mu})\Big].
			\end{align}
			We replace $\log(m_b-\mu) $ in the above expression by using \eqref{log mb-mu} and subsequently 
			with the use of  \eqref{m_b} and \eqref{m_f} in
			the above  expression, this admits the following large $\mu$ expansion given as
			\begin{align}\label{asymp formula bosons}
				\lim_{|\mu \beta | \rightarrow \infty} a_{\cal O}[0, l ] =
				\frac{2^{l-2} \Gamma (l+1)}{\pi  \Gamma (2 l+1)} \mu^{l+1}  \Bigg[ 1+\frac{\sqrt{4\lambda^2+3\pi^2}-2\lambda}{6\lambda}\nonumber \\+ \sum_{n =1}^l \frac{   (l - n +1)_{2n} \zeta(n+1)  }{ 2^n n!} \frac{T^{n+1} }{ |\mu|^{n+1} }+ O(e^{-\beta |\mu|} )\Bigg] .
			\end{align}
			Numerical tests have been performed to check the agreement of this above asymptotic formula against the value of the thermal one point function computed from the expression \eqref{a O bosons}, e.g. we have presented the comparison between the asymptotic formula and the numerical values in the figure \ref{a O l=1} and \ref{a O l=4} for $l=1$ and $l=4$ respectively for different values of $\lambda$. 
			by substituting the numerical solution of the gap equations \eqref{gap eq}.
			Note that we have reinstated the temperature dependence in the last line.
			For spin $l=1 \ {\rm and}\ l=2$, we have
			\begin{align}
				a_{\cal O}[0,1]&=\frac{\left(\sqrt{4 \lambda ^2+3 \pi ^2}+4 \lambda \right) \mu ^2}{24 \pi  \lambda }+\frac{\pi  T^2}{24}+O(e^{-\frac{\mu}{T}}),\\
				a_{\cal O}[0, 2 ] &=\frac{\left(\sqrt{4 \lambda ^2+3 \pi ^2}+4 \lambda \right) \mu ^3}{72 \pi  \lambda }+\frac{1}{24} \pi  \mu  T^2+\frac{T^3 \zeta (3)}{4 \pi }+O(e^{-\frac{\mu}{T}}).
			\end{align}
			Now taking $\lambda\to \infty$ in the above two equations we have 
			\begin{align}
				\lim_{\lambda \to \infty}a_{\cal O}[0,1]&=\frac{\mu ^2}{4 \pi }+\frac{\pi  T^2}{24}+\cdots,\\
				\lim_{\lambda\to \infty} a_{\cal O}[0,2]&=\frac{\mu ^3}{12 \pi }+\frac{1}{24} \pi  \mu  T^2+\frac{T^3 \zeta (3)}{4 \pi }+\cdots.
			\end{align}
			Now these expressions agree with the answers for critical $U(N)$ scalars at large $\mu$ given in equations (2.57) of  \cite{David:2024naf}.
			\subsection*{Fermion bilinears}
\begin{figure}[hbt!]
				
				\begin{subfigure}{.475\linewidth}
					\includegraphics[width=1\linewidth]{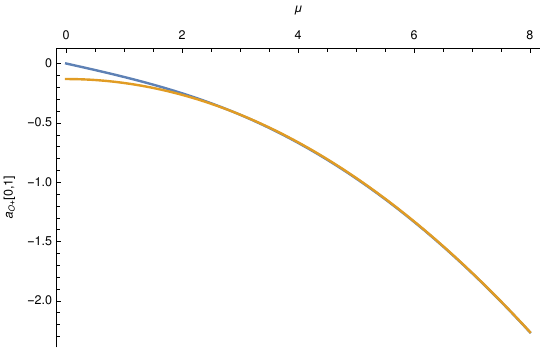}
					\caption{$\lambda=1$}
				\end{subfigure}\hfill 
				\begin{subfigure}{.475\linewidth}
					\includegraphics[width=1\linewidth]{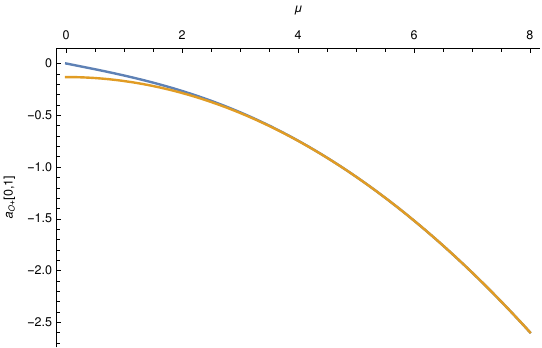}
					\caption{$\lambda=4$}
				\end{subfigure}
				\par\bigskip
				\begin{subfigure}{.475\linewidth}
					\includegraphics[width=1\linewidth]{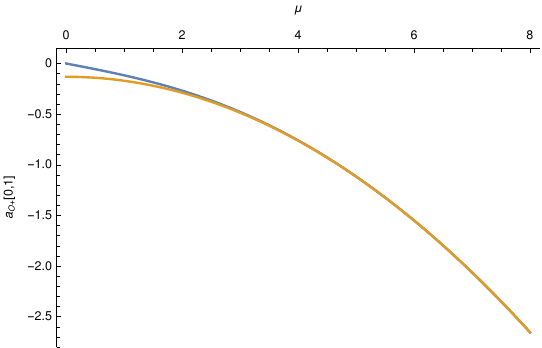}
					\caption{$\lambda=8$}
				\end{subfigure}\hfill 
				\begin{subfigure}{.475\linewidth}
					\includegraphics[width=1\linewidth]{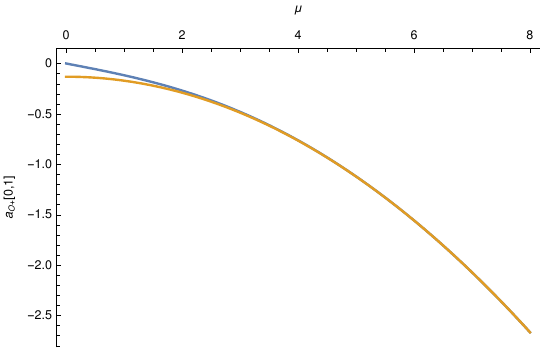}
					\caption{$\lambda=12$}
				\end{subfigure}
				
				\caption{ We compare the asymptotic formula \eqref{asym for fer} for $a_{{\cal O}_+}[0,1]$ against the the numerical value of the $a_{{\cal O}_+}[0,1]$ obtained from the \eqref{fer 1pt fn} with the use numerical solutions of the gap equation \eqref{gap eq} for different values of the coupling constant  $\lambda$. The blue curve denotes the numerical answer while the orange curve stands for the asymptotic formula \eqref{asym for fer}, note that we kept $\beta=1$.}
				\label{a O + l=1}
			\end{figure}

\begin{figure}[t]
				
				\begin{subfigure}{.475\linewidth}
					\includegraphics[width=1\linewidth]{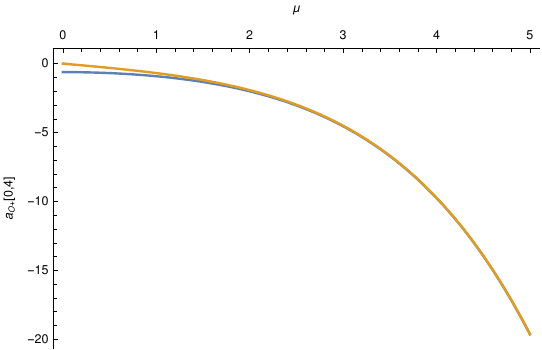}
					\caption{$\lambda=1$}
				\end{subfigure}\hfill 
				\begin{subfigure}{.475\linewidth}
					\includegraphics[width=1\linewidth]{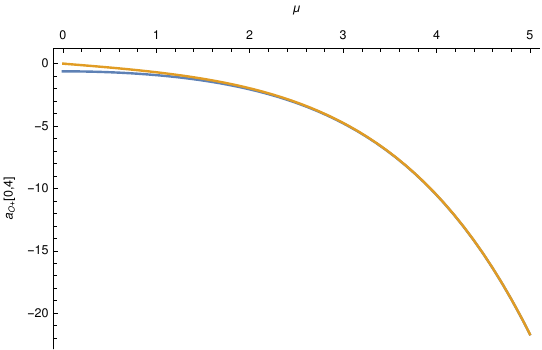}
					\caption{$\lambda=4$}
				\end{subfigure}
				\par\bigskip
				\begin{subfigure}{.475\linewidth}
					\includegraphics[width=1\linewidth]{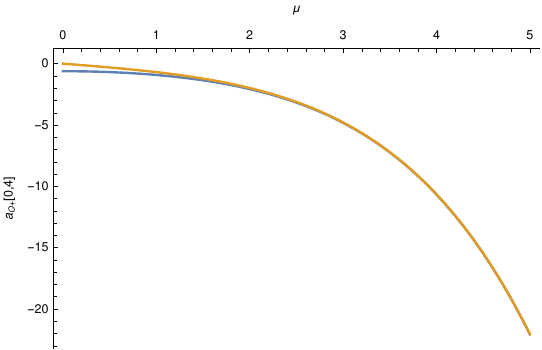}
					\caption{$\lambda=8$}
				\end{subfigure}\hfill 
				\begin{subfigure}{.475\linewidth}
					\includegraphics[width=1\linewidth]{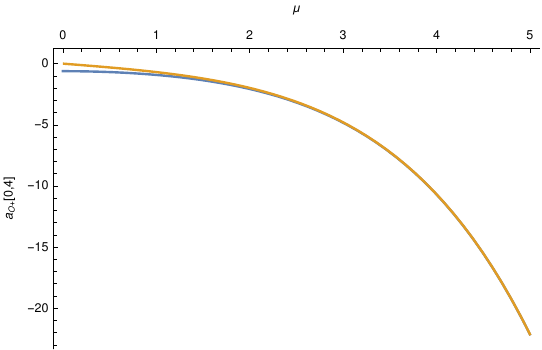}
					\caption{$\lambda=12$}
				\end{subfigure}
				
				\caption{We compare the asymptotic formula \eqref{asym for fer} for $a_{{\cal O}_+}[0,4]$ against the the numerical value of the $a_{{\cal O}_+}[0,4]$ obtained from the \eqref{fer 1pt fn} with the use numerical solutions of the gap equation \eqref{gap eq} for different values of the coupling constant $\lambda$. The blue curve denotes the numerical answer while the orange curve stands for the asymptotic formula \eqref{asym for fer}, note that we kept $\beta=1.$  }
				\label{a O + l=4}
			\end{figure}

			Now we can focus on the thermal one point function of the fermion bilinear operator ${\cal O}_+[0,l] $\footnote{We are only studying ${\cal O}_+[0,l]$ as the thermal expectation of the other two classes of fermion bilinear operators ${\cal O}_-[0,l]$ and ${\cal O}_0[0,l]$ are directly proportional to the thermal expectation of ${\cal O}_+[0,l]$ as shown in \cite{David:2023uya}.} from \eqref{a O+} after analytically continuing to the real chemical potential by $\hat\mu=i \mu$.
			\begin{align}\label{fer 1pt fn}
				a_{\mathcal{\tilde O}_+}[n=0,l]&=\frac{l}{{ \pi \big(\frac{1}{2}\big)_l}}{ \,\sum _{n=0}^l \frac{m_f^{l-n} (l-n+1)_{2 n}}{2^{n+l+2} n!}} \left(\text{Li}_{n+1}\left(-e^{\mu -m_f}\right)+(-1)^l \text{Li}_{n+1}\left(-e^{-m_f-\mu }\right)\right).
			\end{align}
			Now we study the large $\mu$ behaviour of the above expression using the following identity
			\begin{align}
				&	{\rm Li}_s(e^{2\pi iz})+(-1)^s{\rm Li}_s(e^{-2\pi iz})=-\frac{(2\pi i)^s}{s!}B_s(z),\\
				&	{\rm for } \ 0\le {\rm Re} \ z<1\ {\rm if} \ {\rm Im}\ z\ge 0 \ {\rm and} \ 0<{\rm Re}\ z\le 1 \ {\rm if} \ {\rm Im} \ z<0\nonumber.
			\end{align}
			and we obtain
			\begin{align}\label{bernoulli}
				a_{{\cal\tilde O}_+}[0, l] = &-\frac{l m_f^{l} }{ 2^{l+2} \pi  (  \frac{1}{2} )_l}
				\sum_{n=0}^{l } \frac{ (l +1 -n)_{2n}}{ (2 m_f) ^n n!} \frac{(2\pi i)^{n+1}}{(n+1)!} \Big[B_{n+1} \Big(\frac{1}{2}-i\frac{(\mu-m_f)}{2\pi}\Big)+O(e^{-\mu})\Big],
			\end{align}
			$B_n(x)$ denotes the Bernoulli polynomial. Now by substituting the solution of the gap equation for the fermionic thermal mass $m_f$ obtained in \eqref{m_f} with the use of the following notation
			\begin{align}\label{large mu soln}
				m_f=c\mu\qquad {\rm and}\qquad c=\frac{(\sqrt{4 \lambda ^2+3 \pi ^2}-2 \lambda )  }{3 \pi },
			\end{align}
			and using the following identity for the Bernoulli polynomials
			\begin{align}
				B_n(x+y) =\sum_{k=0}^n \binom{n}{k} B_k(x )y ^{n-k},
			\end{align}
			in the equation \eqref{bernoulli} we have
			\begin{align}
				a_{{\cal\tilde O}_+}[0,l]=-\frac{l m_f^{l} }{ 2^{l+2} \pi  (  \frac{1}{2} )_l}
				\sum_{n=0}^l \sum_{k=0}^{n+1} (2\pi i)^k
				\frac{(l+1-n)_{2 n}}{2^n n! (n+1)! }\binom{n+1}{k} B_k\Big(\frac{1}{2}\Big) \frac{(1-c)^{n+1-k} \mu^{1-k}}{c^n}.
			\end{align}
			Note that we have neglected the $O(e^{-\mu})$ correction from  \eqref{bernoulli}.
			Now by interchanging the order of the sum over $n$ and $k$ in the above expression we can organise $ a_{{\cal\tilde O}_+}[0,l]$ systematically in orders of $\mu$ at large $\mu$ as given below
			\begin{align}\label{asym for fer}
				a_{{\cal\tilde O}_+}[0,l]=	-\frac{l (c \mu )^l }{\pi  2^{l+2} \left(\frac{1}{2}\right)_l}
				\sum _{k=0}^{l+1} (2 \pi i  )^{k} B_k\left(\frac{1}{2}\right) \mu ^{1-k}
				\sum _{n=k-1 }^l &\frac{ \binom{n+1}{k} (1-c)^{n+1-k} (l-n+1)_{2 n}}{2^n n! (n+1)! c^n}	.
			\end{align}
			The sum over $n $ can be performed and expressed in terms of the regularised hypergeometric function $_2\tilde F_1$ as given below
			\begin{align}
				a_{{\cal\tilde O}_+}[0,l]=	-\frac{l (c \mu )^l }{\pi  2^{l+2} }
				\sum _{k=0}^{l+1} \frac{(2 \pi i  )^{k}B_k\left(\frac{1}{2}\right) }{\left(\frac{1}{2}\right)_l(2c\mu)^{k-1}} 
			\frac{ \, _2\tilde{F}_1\left(k-l-1,k+l;k;\frac{c-1}{2 c}\right)}{k! (k+l)_{2-2k}}
			\end{align}
			We have tested this asymptotic formula against the values of the thermal one point function \eqref{fer 1pt fn} obtained using the numerical solution of the gap equations \eqref{gap eq}, e.g., we show the comparison between the asymptotic answer given above and the numerically obtained value  in the figure \ref{a O + l=1} and \ref{a O + l=4} for $l=1$ and $l=4$ respectively for different values of $\lambda$.
			For illustration, from the above formula we explicitly write down the large $\mu$ asymptotic formula for  thermal expectation values of the spin-1 and spin-2 current constructed from fermion bilinear as given below after reinstating the temperature dependence 
			\begin{align}\label{al+=2}
				a_{{\cal\tilde O}_+}[0,2]=-\frac{\left(3\pi^2-4 \lambda ^2+2 \lambda\sqrt{4 \lambda ^2+3 \pi ^2}  \right) \mu ^3}{54 \pi ^3}-\frac{\pi  \mu T^2 }{12}+\cdots,
			\end{align}
			and
			\begin{align}\label{al+=1}
				a_{{\cal\tilde O}_+}[0,1]=-\frac{\left(2 \lambda  \left(\sqrt{4 \lambda ^2+3 \pi ^2}-2 \lambda \right)+3 \pi ^2\right) \mu ^2}{36 \pi ^3}-\frac{\pi T^2 }{24}+\cdots.
			\end{align}
			
			We know that at large $\lambda$ these answers should agree with the one point functions of free fermions at large $\mu$ as mentioned earlier, we have verified that it is indeed true.
			Note that the large $\mu$ asymptotic form of the higher spin currents for both scalars and fermions  given in \eqref{asymp formula bosons} and \eqref{asym for fer} admits branch cut singularity at $\lambda=\pm \frac{\sqrt{3}}{2}\pi i $ in the analytically continued complex $\lambda $ plane. 
			For imaginary values of $\lambda$ the theory lies on the non-unitary domain.

				\section{Agreement with the stress tensor and spin-1 current} \label{sec 6}
				In this section we discuss the agreement of the stress tensor and spin-1 current evaluated from the partition function with the correct combination of the bosonic and fermionic spin-2 currents and spin-1 currents respectively obtained from the OPE inversion formula. 
				The partition function at large $N$ given in \eqref{2 integrals} is evaluated at the saddle point is given by
				\begin{align}\label{pat fn Z}
					\log Z&= -N\beta V(\frac{m_f^3}{8\lambda}-\frac{m_b^2m_f}{8\lambda}-\frac{1}{\beta}\log Z_b(m_b,-i\hat\mu)-\frac{1}{\beta}\log Z_f(m_f,-i\hat\mu)).
				\end{align}
				After substituting the expression for $\log Z_b(m_b)$ and $\log Z_f(m_f)$ from \eqref{Z_b} and \eqref{Z_f} respectively in \eqref{pat fn Z}, we use the gap equations \eqref{gap eq} to obtain $\log Z$ to be
				\begin{align}\label{log Z}
					\log Z=-\frac{N V  }{6\pi \beta^2} \Big[m_b^2\beta^2(\log(1-e^{-m_b\beta+i\hat \mu\beta})+\log(1-e^{-m_b\beta-i\hat \mu\beta}))\nonumber\\
					+3 m_b \beta ({\rm Li}_2(e^{-m_b\beta+i\hat \mu\beta})+{\rm Li}_2(e^{-\beta m_b-i\hat \mu\beta}))\nonumber\\
					+	{\rm Li}_3(e^{-\beta m_b\beta+i\hat \mu\beta})+{\rm Li}_3(e^{-\beta m_b\beta-i\hat \mu\beta})\nonumber\\
					-m_f^2\beta^2(\log(1+e^{-m_f\beta+i\hat \mu\beta})+\log(1+e^{-m_f\beta-i\hat \mu\beta}))\nonumber\\
					-3 m_f \beta ({\rm Li}_2(-e^{-m_f\beta+i\hat \mu\beta})+{\rm Li}_2(-e^{-\beta m_f-i\hat \mu\beta}))\nonumber\\
					-	{\rm Li}_3(-e^{-\beta m_f\beta+i\hat \mu\beta})-{\rm Li}_3(-e^{-\beta m_f\beta-i\hat \mu\beta})
					\Big].
				\end{align}
				\subsection*{Stress tensor}
				The stress tensor can be evaluated by differentiating  the partition function \eqref{pat fn Z} in the following manner
				\begin{align}\label{T 00}
					T_{00}&=\partial_\beta \log Z |_{\hat \mu\beta={\rm constant}} \nonumber,\\
					&=-\frac{m_f^3}{8\lambda}+\frac{m_b^2m_f}{8\lambda}\nonumber
					+\frac{1}{6\pi \beta ^3}\Big[
					\beta ^2 m_b^2 (3 \log (1-e^{-\beta  m_b+i \hat \mu\beta })+3 \log (1-e^{-\beta m_b-i \hat \mu\beta })+\beta  m_b)\nonumber\\
					&-6 \beta  m_b (\text{Li}_2(e^{i \hat \mu\beta -m_b \beta })+\text{Li}_2(e^{-m_b \beta -i \hat \mu\beta }))-6 \text{Li}_3(e^{i \hat \mu\beta -m_b \beta })-6 \text{Li}_3(e^{-m_b \beta -i \hat \mu\beta })\nonumber\\
					&
					-\beta ^2 m_f^2 (3 \log (1+e^{-\beta  m_f+i \hat \mu\beta })+3 \log (1+e^{-m_f\beta-i \hat \mu\beta })+\beta  m_f)\nonumber\\
					&	+6 \beta  m_f (\text{Li}_2(-e^{i \hat \mu\beta -m_f \beta })+\text{Li}_2(-e^{-m_f \beta -i \hat \mu\beta }))+6 \text{Li}_3(-e^{i \hat \mu\beta -m_f \beta })+6 \text{Li}_3(-e^{-m_f \beta -i \hat \mu\beta })\Big].
				\end{align}
				Now with the use of the gap equations \eqref{gap eq} in \eqref{T 00} we obtain the stress tensor at the conformal fixed point to be
				\begin{align}\label{T 00 1}
					T_{00}
					&=\frac{1}{3\pi \beta ^3}\Big[
					\beta ^2 m_b^2 (\log (1-e^{-\beta  m_b+i \hat \mu\beta })+3 \log (1-e^{-\beta m_b-i \hat \mu\beta }))\nonumber\\
					&-3 \beta  m_b (\text{Li}_2(e^{i \hat \mu\beta -m_b \beta })+\text{Li}_2(e^{-m_b \beta -i \hat \mu\beta }))-3 \text{Li}_3(e^{i \hat \mu\beta -m_b \beta })-3 \text{Li}_3(e^{-m_b \beta -i \hat \mu\beta })\nonumber\\
					&
					-\beta ^2 m_f^2 (\log (1+e^{-\beta  m_f+i \hat \mu\beta })+3 \log (1+e^{-m_f\beta-i \hat \mu\beta }))\nonumber\\
					&	+3 \beta  m_f (\text{Li}_2(-e^{i \hat \mu\beta -m_f \beta })+\text{Li}_2(-e^{-m_f \beta -i \hat \mu\beta }))+3 \text{Li}_3(-e^{i \hat \mu\beta -m_f \beta })+3 \text{Li}_3(-e^{-m_f \beta -i \hat \mu\beta })\Big],\nonumber\\
					&\equiv T_{00}^b+T_{00}^f.
				\end{align}
				$T_{00}^b$ and $T_{00}^f$ are contributions to the total stress tensor from the scalars and fermions in the action respectively. The terms in \eqref{T 00 1} involving $m_b$ are contained in $T_{00}^b$ and the rest is contained in $T_{00}^f$.\\
				
				Comparing \eqref{T 00 1} and \eqref{log Z} we have, the expected relation between stress tensor and the free energy of a thermal CFT on $S^1\times R^{2}$
				\begin{align}
					T_{00}=- \frac{2 \log Z}{N\beta}.
				\end{align}
				Thus the above expression \eqref{T 00 1} is the stress tensor evaluated from the partition function. Now to obtain the stress tensor of the theory from the thermal one point functions evaluated using OPE inversion we should add the scalar spin-2 current ${\cal O}[0,2]$ from \eqref{a O} and fermion spin-2 current ${\cal O}_+[0,2]$ from \eqref{a O+} respectively with the correct normalisations. We use the normalisations such that the bosonic spin-2 currents ${\cal O}[0,2]$  exactly match with the stress tensor for a free complex scalar and fermion spin-2 current ${\cal O}_+[0,2]$ matches with the stress tensor for a free fermion when $m_b=m_f=\mu=0$.
				\begin{align}\label{-4+2}
					&-4a_\mathcal{O}[0,2]+ 2a_{\mathcal{\tilde O}_+}[0,2]=\frac{1}{3 \pi}\Big[m_b^2 (\log (1-e^{i\hat \mu -m_b})+\log (1-e^{-i\hat \mu -m_b}))\nonumber\\
					& \qquad \qquad
					-3 m_b (\text{Li}_2(e^{-m_b+i\hat \mu })+\text{Li}_2(e^{-i\hat \mu -m_b}))-3 (\text{Li}_3(e^{-m_b+i\hat \mu })+\text{Li}_3(e^{-i\hat \mu -m_b}))\nonumber\\
					& \qquad\qquad
					- m_f^2 (\log (e^{i\hat \mu -m_f}+1)+\log (e^{-i\hat \mu -m_f}+1))\nonumber\\
					&
					+3 m_f (\text{Li}_2(-e^{-m_f+i\hat \mu })+\text{Li}_2(-e^{-i\hat \mu -m_f}))+3 (\text{Li}_3(-e^{-m_f+i\hat \mu })+\text{Li}_3(-e^{-i\hat \mu -m_f}))
					\Big].
				\end{align}
				To elaborate the scheme of normalisations we use the following  answers for the  stress tensor of a free massless complex scalar and a fermion respectively at finite temperature in $d=3$
				\begin{align}
					T_{00}^{\rm free\ boson}=-\frac{2\zeta(3)}{\pi \beta^3}, \qquad T_{00} ^{\rm free\ fermion}=-\frac{3 \zeta (3)}{2 \pi  \beta ^3}.
				\end{align}
				Now it is easy to see from equation \eqref{a O}, by taking the free theory limit $m_b=0$ and $\mu=0$, with $\beta=1$.
				\begin{align}
					-4 a_{\cal O}[0,2]|_{m_b=0,\mu=0}=T_{00}^{\rm free\ boson}.
				\end{align}
				Similarly  by taking $m_f=\mu=0$ in the equation \eqref{a O+} we have
				\begin{align}
					2 a_{{\cal\tilde O}_+}[0,2]|_{m_f=0,\mu=0}=T_{00} ^{\rm free\ fermion}.
				\end{align}
				Thus the combination $-4a_\mathcal{O}[0,2]+ 2a_{\mathcal{O}_+}[0,2] $ agrees with the stress tensor for our model at the free CFT point where the coupling constant $\lambda=0$.
				We can also see that the expression \eqref{-4+2} exactly matches with the stress tensor of the theory \eqref{T 00 1} at the interacting CFT obtained from the direct calculation of the partition function at $\beta=1$. 
				\begin{align}
					T_{00}=-4a_\mathcal{O}[0,2]+ 2a_{\mathcal{\tilde O}_+}[0,2].
				\end{align}It clearly  shows that the OPE coefficients $C_{\phi\phi {\cal O}[0,2]}$  and $C_{\psi\psi {\cal O}_+[0,2]} $ at the interacting CFT with any value of $\lambda$ do not alter from that at the free CFT point which was argued during the discussion of the the OPE inversion formula. 
				
				\subsection*{Spin-1 Current}
				We evaluate the spin-1 current in the theory from the partition function \eqref{pat fn Z} by differentiating w.r.t $\hat\mu$ as given below
				\begin{align}\label{J}
					&J=\frac{i}{\beta}\partial_{\hat \mu} \log Z,\\\nonumber
					&=\frac{1}{{2 \pi  \beta ^2}}\Big[\beta  m_b (\log (1-e^{-\beta  m_b+i \beta  \hat \mu })-\log (1-e^{-\beta  (m_b+i \hat \mu )}))\nonumber\\
					&+\text{Li}_2(e^{-\beta  (m_b+i \hat \mu )})-\text{Li}_2(e^{i \beta  \hat \mu -m_b \beta })\nonumber
					+\beta  m_f (\log (1+e^{-\beta  (m_f+i \hat \mu )})-\log (1+e^{-\beta  m_f+i \beta  \hat \mu }))\nonumber\\
					&-\text{Li}_2(-e^{-\beta  (m_f+i \hat \mu )})+\text{Li}_2(-e^{i \beta  \hat \mu -m_f \beta })\Big]\nonumber,\\
					&\equiv J^b+J^f\nonumber.
				\end{align}
				The contribution to the total spin-1 current in the theory due to the scalars is given by $J_b$ containing the terms involving $m_b$. While $J_f $ contains the rest of the terms involving $m_f$ which give the contribution due to fermions.
				The spin-1 current from the OPE inversion formula is obtained by adding thermal one point function of scalar spin-1 current ${\cal O}[0,1]$ from \eqref{a O} and fermion spin-1 current ${\cal O}_+[0,1] $ from the equation \eqref{a O+}
				\begin{align}\label{spin 1}
					a_{\cal O}[0,1]-a_{{\cal\tilde O}_+}[0,1]=\frac{1}{4\pi} \Big(m_b \log (1-e^{i\hat\mu -m_b})-m_b \log (1-e^{-i\hat\mu -m_b})\nonumber\\
					-\text{Li}_2(e^{-m_b+i\hat\mu })+\text{Li}_2(e^{-i\hat\mu -m_b})+m_f \log (e^{-i\hat\mu -m_f}+1)-m_f \log (e^{i\hat\mu -m_f}+1)\nonumber\\
					-\text{Li}_2(-e^{-m_f-i\hat\mu })+\text{Li}_2(-e^{i\hat\mu -m_f})\Big),
				\end{align}
				with $\beta=1$  comparing \eqref{J} and \eqref{spin 1} we have
				\begin{align}
					a_{\cal O}[0,1]-a_{{\cal\tilde O}_+}[0,1]=\frac{J}{2}.
				\end{align}
				This is true for all values $m_b,m_f$ and $\mu$, or in other words it is true for the Gaussian fixed point as well as the interacting CFT which shows that the OPE coefficient $C_{\phi\phi{\cal O}[0,1]}$ and $C_{\psi\psi{\cal\tilde O}_+[0,1]}$ keep fixed across the free CFT as well as the interacting CFT with arbitrary values of coupling.
				\subsection*{Vanishing of $C_{\phi\phi {\cal\tilde O}_+[0,1]},C_{\phi\phi {\cal\tilde O}_+[0,2]} $ }
				Here we show that the assumption $C_{\phi\phi {\cal\tilde O}[0,l]}=0$ used just before \eqref{a O}
				is indeed true for $l=1$ and $l=2$. Thus  we proceed without the assumption that $C_{\phi\phi{\cal\tilde O}_+[0,l]}=0$, from the equation \eqref{hat a} and \eqref{ph ph OPE} we have
				\begin{align}
					&	(a^b_{\mathcal{O}}[0,l]+ a^b_{{\cal\tilde O}_+}[0,l])\\
					&=\frac{K_l}{ l!}\sum_{n=0}^l\frac{ m_b^{l-n} (l-n+1)_{2 n} }{2^{l+n} n! }\times\Big(\text{Li}_{n+1}(e^{-m_b-i \hat \mu })+(-1)^l \text{Li}_{n+1}(e^{-m_b+i \hat \mu })\Big) ,
					\qquad	{\rm for }\  l>0.\nonumber
				\end{align}
				For $l=1$ we have \begin{align}\label{l=1}
					(a^b_{\mathcal{O}}[0,1]+ a^b_{{\cal\tilde O}_+}[0,1])=\frac{1}{4\pi}	\Big(m_b \left(\log (1-e^{i\hat\mu -m_b})-\log (1-e^{-i\hat\mu -m_b})\right)\nonumber\\-\text{Li}_2(e^{-m_b+i\hat\mu })+\text{Li}_2(e^{-i\hat\mu -m_b})\Big),
				\end{align}
				For $l=2$
				\begin{align}\label{l=2}
					(a^b_{\mathcal{O}}[0,2]+ a^b_{{\cal\tilde O}_+}[0,2])=	\frac{1}{{12 \pi }} \Big(-m_b^2 (\log (1-e^{-i\hat\mu -m_b})+\log (1-e^{i\hat\mu -m_b})) \nonumber\\+3 m_b (\text{Li}_2(e^{-m_b-i\hat\mu })+\text{Li}_2(e^{i\hat\mu -m_b}))+3 (\text{Li}_3(e^{-m_b-i\hat\mu })+\text{Li}_3(e^{i\hat\mu -m_b}))\Big).
				\end{align}
				Now using the result from the partition function given in \eqref{J}
				\begin{align}\label{J_b}
					J_b=\frac{1}{2\pi}	\Big(m_b \left(\log (1-e^{i\hat\mu -m_b})-\log (1-e^{-i\hat\mu -m_b})\right)-\text{Li}_2(e^{-m_b+i\hat\mu })+\text{Li}_2(e^{-i\hat\mu -m_b})\Big),
				\end{align}and from \eqref{T 00 1}
				\begin{align}\label{T 00 b}
					T_{00}^b&=-\frac{1}{{3 \pi }} \Big(-m_b^2 (\log (1-e^{-i\hat\mu -m_b})+\log (1-e^{i\hat\mu -m_b})) \nonumber\\
					&+3 m_b (\text{Li}_2(e^{-m_b-i\hat\mu })+\text{Li}_2(e^{i\hat\mu -m_b}))+3 (\text{Li}_3(e^{-m_b-i\hat\mu })+\text{Li}_3(e^{i\hat\mu -m_b}))\Big),
				\end{align}
				As our starting point was that $a_{\cal O}^b[0,l]$ gives the thermal expectation of the spin-$l$ current constructed from the scalar bilinears which for $l=1$ and $2$ is also evaluated from the partition function as $J_b$ and $T_{00}^b$ as given in above two equations. Thus comparing \eqref{l=1} with \eqref{J_b} and \eqref{l=2} with \eqref{T 00 b}, it is consistent with the fact that 
				\begin{align}
					a_{\cal O}^b[0,1]=2 J_b \qquad {\rm and} \qquad a_{{\cal \tilde O}_+}^b[0,1]=0,\\
					a_{\cal O}^b[0,2]=-4 T_{00}^b \qquad {\rm and} \qquad a_{{\cal \tilde O}_+}^b[0,2]=0.
				\end{align}
				And as $\langle {\cal \tilde O}_+[0,l]\rangle\ne 0$ we  have
				\begin{align}
					C_{\phi\phi{\cal\tilde O}_+[0,1]}=C_{\phi\phi{\cal\tilde O}_+[0,2]}=0.
				\end{align}
				
					\section{Conclusions}
					
					We have studied the large $N$ Wess-Zumino model in presence of the chemical potential($\mu$) in $3d$. Starting from the supersymmetric Wess-Zumino model for $2N$ real scalars and $2N$ real fermions without chemical potential in the fundamental of $O(2N) $, we construct the $U(N)$ Wess-Zumino model with $N$ complex scalars and $N$ Dirac fermions. We have introduced a chemical potential in this setup at finite temperature. The non-trivial fixed point of the model is obtained at the saddle point of the action in the zero modes of the auxiliary fields used to linearise the action at large $N$. The saddle point condition is known as the gap equation relating the thermal masses to the chemical potential $\mu$ at finite temperature. These thermal masses, obtained by solving the gap equations, describe the non-trivial fixed point of the theory. The interaction between boson and fermion degrees of freedom in this model leads to a pair of coupled gap equations in this case.  \\
					
					We have found the analytically solvable regime of the gap equations for the large $N$ Wess-Zumino model in $3d$  at large chemical potential($\mu$) for arbitrary coupling($\lambda$).  By using these analytic solutions of the gap equation at large $\mu$, we evaluate the partition function, stress tensor and spin-1 current of the theory as a perturbative expansion at large $\mu$. 					
					 Applying the OPE inversion formula we also evaluate the higher spin currents of the theory as a perturbative expansion at large $\mu$. As a direct consequence of the study of the OPE inversion formula we find that at the leading order in large $N$ the OPE coefficients $C_{\phi\phi {\cal O}_0[0,l]}$ and $C_{\psi\psi{\cal O}[0,l]}$ vanish for all values of $l$ and thus the relevant three point functions also vanish. Although our OPE analysis can not directly conclude whether the coefficients $C_{\phi\phi{\cal O}_+[0,l]}$ and $C_{\phi\phi{\cal O}_-[0,l]}$ also vanish. But comparing the results from the OPE inversion formula and the results obtained from the partition function we have shown that the $C_{\phi\phi{\cal O}_+[0,l]}$ indeed vanishes for $l=1,2$.  It would be interesting to evaluate these coefficients directly.\\
					 
					  An interesting problem would be to extend this study on curved backgrounds such as $S^1\times S^2$, the method developed in \cite{David:2024pir} for $O(N)$ model in this geometry can be useful in this context. The large chemical potential limit may serve as a solvable limit in various other related models. The analysis presented in this paper perhaps contain useful ingredients necessary for the study of the large $N$ Gross-Neveu-Yukawa  model \cite{Moshe:2003xn} at finite chemical potential and finite temperature.\\
					  
					  In the study of thermal one point functions of higher spin currents, large spin behaviour of these one point functions emerges as an interesting observation\cite{Iliesiu:2018fao,David:2023uya,David:2024naf,David:2024pir}. For the  large $N$ model of complex scalars and the Gross-Neveu model at finite chemical potential, the large spin limit of the 1-point functions of the currents at the non-trivial fixed point resembles the currents in free CFT. The dependence on the chemical potential($\mu$) reduces to a simple factor of $\sinh\mu$ or $\cosh\mu$ \cite{David:2024naf} for odd or even spin respectively.  Thus an interesting exploration would be to understand the interplay between large spin($l$) and large $\mu$ and to quantify it.\\
					  
					   The study of the class of vector models at finite $N$ should be important for practical interests. For the $O(N) $ model at finite temperature and zero chemical potential, a bootstrap method \cite{Barrat:2024fwq} is developed based on a set of sum rules \cite{Marchetto:2023xap}. Using this method, thermal OPE coefficients are computed at finite $N$. It would be useful to generalise such a method in the presence of a finite chemical potential. The generalisation of the techniques of the OPE inversion formula for the large $N$ Chern-Simons matter theories \cite{Giombi:2011kc,Aharony:2011jz} still remains an interesting open question.

					\section*{Acknowledgments}
					I am deeply indebted to Justin David for numerous discussions, which played a crucial role in shaping this work, as well as for his valuable comments on the final draft. I would like to thank Raghu Mahajan, Ashoke Sen and the String theory group at ICTS-TIFR Bengaluru for their insightful comments during my talk there based on my previous works. I would like to thank the organisers and participants of the National String Meeting 2024, IIT Ropar, for providing me the opportunity to present  my previous work and attend encouraging sessions on related topics.
					
					\appendix
					\section{Partition function in $3d$ for free boson and fermions}
					\label{A}
					In this appendix we show the calculation of the partition function for free bosons and fermions in $d=3$ to obtain the results used in \eqref{Z_b} and \eqref{Z_f}. We will elaborate on the analytic continuation used to obtain these results.
					
					\subsection*{Free complex scalars}
					We consider the Euclidean action for a complex scalar of mass $m$ at presence of chemical potential $\mu=-i\hat \mu$ as given below
					\begin{align}
						S=\int d^3x ((D^\nu\tilde\phi)^*D_\nu\tilde\phi+m^2\tilde\phi^*\tilde\phi).
					\end{align}
					As defined earlier $D_\nu=\partial_\nu-i\hat \mu \delta_{\nu 0} $. The partition function of this action given by the following Euclidean path integral
					\begin{align}
						&Z_b=\int {\cal D}\tilde \phi^*{\cal D}\tilde\phi e^{-S},\qquad {\rm with} \nonumber\\
						& \tilde\phi(\tau+\beta,\vec x)=\tilde\phi(\tau,\vec x),\ \tilde\phi^*(\tau+\beta,\vec x)=\tilde\phi^*(\tau,\vec x).
					\end{align}
					The path integral is performed in the standard approach by the use of mode expansion of the free  field $\phi(x)$ to give the partition function as 
					\begin{align}
						\log Z_b=-  \sum_{n =-\infty}^\infty \int \frac{d^{2} p}{(2\pi)^{2}}  \log
						\left[ \frac{  (2\pi n-\hat \mu\beta)^2 }{\beta^2}  + \vec p^2 +m^2 \right].
					\end{align}
					Now the sum over Matsubara frequencies can be performed using the following formula for the regulated sum \cite{Klebanov:2011uf}
					\begin{align}\label{regulated sum cos}
						\sum_{n=-\infty}^\infty \log \Big(\frac{(n+\alpha)^2}{q^2}+a^2\Big)=\log [2\cosh (2\pi q|a|)-2\cos(2\pi\alpha)].
					\end{align}
					to obtain
					\begin{align}\label{Z free boson}
						\log {Z}_b&=-\frac{1}{\beta^{2}}\int \frac{d^2k}{(2\pi)^2}\Big[\sqrt{\vec{k}^2+m_{\rm th}^2\beta^2}+2 {\rm Re}\Big(\log\big[1-e^{-\sqrt{\vec k^2+m_{\rm th}^2\beta^2}+i\beta\hat \mu}\big]\Big)\Big].
					\end{align}
					The 1st term from the above expression is a diverging integral but we regularise it with the use of the analytic continuation of the following standard integral formula as was done in \cite{Giombi:2019upv},
					\begin{align}\label{analy cont}
						-\frac{1}{\beta^{2}}\int \frac{dk k}{(2\pi)}({k}^2+m_{\rm th}^2\beta^2)^{-\alpha}\Big|_{\alpha=-\frac{1}{2}}=-\frac{\left(\beta ^2 m^2\right)^{1-\alpha }}{4 \pi  (\alpha -1) \beta ^2}.
					\end{align}
					The integral in the second term from \eqref{Z free boson} is convergent and can be evaluated using the expansion of the $\log (1+x)$ in small $x$ and resumed back to give the following result
					\begin{align}
						\log Z_b=\frac{\beta  m^3}{6 \pi }	+{\rm Re}\frac{\beta  m \text{Li}_2\left(e^{-\beta  (m-i \mu )}\right)+\text{Li}_3\left(e^{-\beta  (m-i \mu )}\right)}{\pi  \beta ^2}.
					\end{align}
					\subsection*{Free fermions}
					Considering the Euclidean action for a free massive fermion  in $d=3$ with chemical potential $\mu=-i\hat \mu$ as given below
					\begin{align}
						S=\int d^3x (\tilde\psi^\dagger\gamma^\mu D_\mu \tilde\psi-m\tilde\psi^\dagger\tilde\psi).
					\end{align}
					The partition function of this action is given by the following Euclidean path integral
					\begin{align}
						Z_b&=\int {\cal D} \tilde\psi^\dagger{\cal D}\tilde\psi e^{-S} ,\qquad {\rm with}\nonumber\\
						\qquad \tilde\psi(\tau+\beta,\vec x)&=-\tilde\psi(\tau,\vec x),\ \tilde\psi^\dagger(\tau+\beta,\vec x)=-\tilde\psi^\dagger(\tau,\vec x).
					\end{align}
					And it is evaluated as
					\begin{align}
						\log Z_f=\sum_{n =-\infty}^\infty \int \frac{d^{2} p}{(2\pi)^{2}}  \log
						\left[ \frac{  ((2n+1)\pi -\hat \mu\beta)^2 }{\beta^2}  + \vec p^2 +m^2 \right].
					\end{align}
					Performing the Matsubara sum using \eqref{regulated sum cos} we get
					\begin{align}
						\log Z_f =  \frac{ 1}{\beta^{2k}}  \int \frac{d^{2} p}{(2\pi)^{2}} 
						\left[  \sqrt{ \vec p^2 + m^2 \beta} +  2 {\rm Re}\log \Big(1+ e^{ - \sqrt{ \vec p^2 + m^2 \beta^2 } +i\hat \mu\beta}  \big)
						\right].
					\end{align}
					Again using the analytic continuation of the integral \eqref{analy cont} we evaluate the above integral to obtain
					\begin{align}
						\log Z_f&=-\frac{1}{6 \pi  \beta ^2}\Big(\beta ^3 m_f^3+3 \beta  m_f (\text{Li}_2(-e^{-\beta  (m_f+i\hat \mu )})+\text{Li}_2(-e^{-\beta  (m_f-i \hat\mu )}))\nonumber\\
						&\qquad+3 \text{Li}_3(-e^{-\beta  (m_f+i \hat\mu )})+3 \text{Li}_3(-e^{-\beta  (m_f-i \hat\mu )})\Big).
					\end{align}
					\bibliographystyle{JHEP}
					\bibliography{ref} 	
				\end{document}